\documentclass[journal,12pt,draftclsnofoot,onecolumn]{IEEEtran}
\usepackage{graphics}
\usepackage{graphicx}
\usepackage{epsfig}
\usepackage{amsmath,amssymb}
\usepackage{mathrsfs,amsfonts,fancyhdr}
\usepackage{multirow}
\usepackage{multicol}
\usepackage{enumitem}
\usepackage{cite}
\usepackage{amsthm}
\usepackage{epstopdf}
\usepackage{hhline}
\newtheorem{thm}{Theorem}[]
\newtheorem{cor}{Corollary}[]

\theoremstyle{remark}
\newtheorem{rem}{Remark}[]
\newcommand\norm[1]{\left\lVert#1\right\rVert}

\allowdisplaybreaks
\begin{document}

\title{{\LARGE Resource Allocation for Massive MIMO HetNets with Quantize-Forward Relaying}}
\author{\normalsize{Ahmad Abu Al Haija, Min Dong, Ben Liang,  Gary Boudreau}
\thanks{Ahmad Abu Al Haija was with the University of Toronto (email: ahmad.abualhaija@utoronto.ca), Ben Liang is with the University of Toronto (email: liang@ece.utoronto.ca),
Min Dong is with Ontario Tech University (email: Min.Dong@ontariotechu.ca),
and Gary Boudreau is with Ericsson Canada (email: gary.boudreau@ericsson.com).}
}
\maketitle
\vspace{-20mm}
\begin{abstract}
\renewcommand{\arraystretch}{0.8}
We investigate  how massive MIMO impacts the uplink transmission design in a heterogeneous network (HetNet) where multiple users communicate with a macro-cell base station (MCBS) with the help of a small-cell BS (SCBS) with zero-forcing (ZF) detection at each BS.
We first analyze the quantize-forward (QF) relaying scheme with joint decoding (JD) at the MCBS. To maximize the rate region, we optimize the quantization of all  user data streams at the SCBS by developing a novel water-filling algorithm that is based on the Descartes' rule of signs.  Our result shows that as a user link to the SCBS becomes stronger than that to the MCBS, the SCBS deploys finer quantization to that user data stream.
We further propose a new simplified scheme through Wyner-Ziv (WZ) binning and time-division (TD) transmission at the SCBS,  which allows not only sequential but also separate decoding of each user message at
the MCBS. For this new QF-WZTD scheme, the optimal quantization parameters are identical to that of  the QF-JD scheme while the phase durations are conveniently optimized as functions of the quantization parameters. Despite its simplicity, the QF-WZTD scheme achieves the same rate performance of the QF-JD scheme, making it an attractive option for future HetNets.
\end{abstract}

\IEEEpeerreviewmaketitle
\vspace{-4mm}
\section{Introduction}\label{sec:intro}
The development of fifth-generation (5G) and beyond cellular networks  aims to drastically improve the data rate of current networks to meet the staggering pace of increasing demand and serve a large number of connected devices. To achieve these goals, several key technologies  are proposed, including massive MIMO systems, heterogeneous networks (HetNets) with wireless backhaul and full-duplex transmission \cite{5GRD}. To improve the performance and reduce the complexity, it is important that these technologies are sufficiently designed and utilized together.
%
\vspace{-4mm}
\subsection{Background and Motivation}
Consider the HetNet in Fig. \ref{fig:HETNET} where there are $K$ user-equipment (UE) inside the small cell  served by the base station (BS).   For this HetNet, the uplink transmission is  theoretically modeled   as a multiple access relay channel (MARC)  \cite{MARC1}  where each UE resembles  a source $({\cal S})$, the SCBS  resembles the relay $({\cal R})$ and the MCBS resembles  the destination $({\cal D})$. Fig. \ref{fig:system_model} shows the MARC with two sources. In the HetNet uplink transmission, the $K$ UEs  transmit their information to the MCBS with the help of the SCBS at rates $R_1,$ $R_2,$ $\ldots,$ and $R_K$. To improve the rates,  this paper aims to design a simple yet efficient transmission scheme that exploits  the following technologies:
\noindent
\begin{figure}[!t]
\begin{minipage}[b]{0.48\linewidth}
    \includegraphics[width=0.9\textwidth]{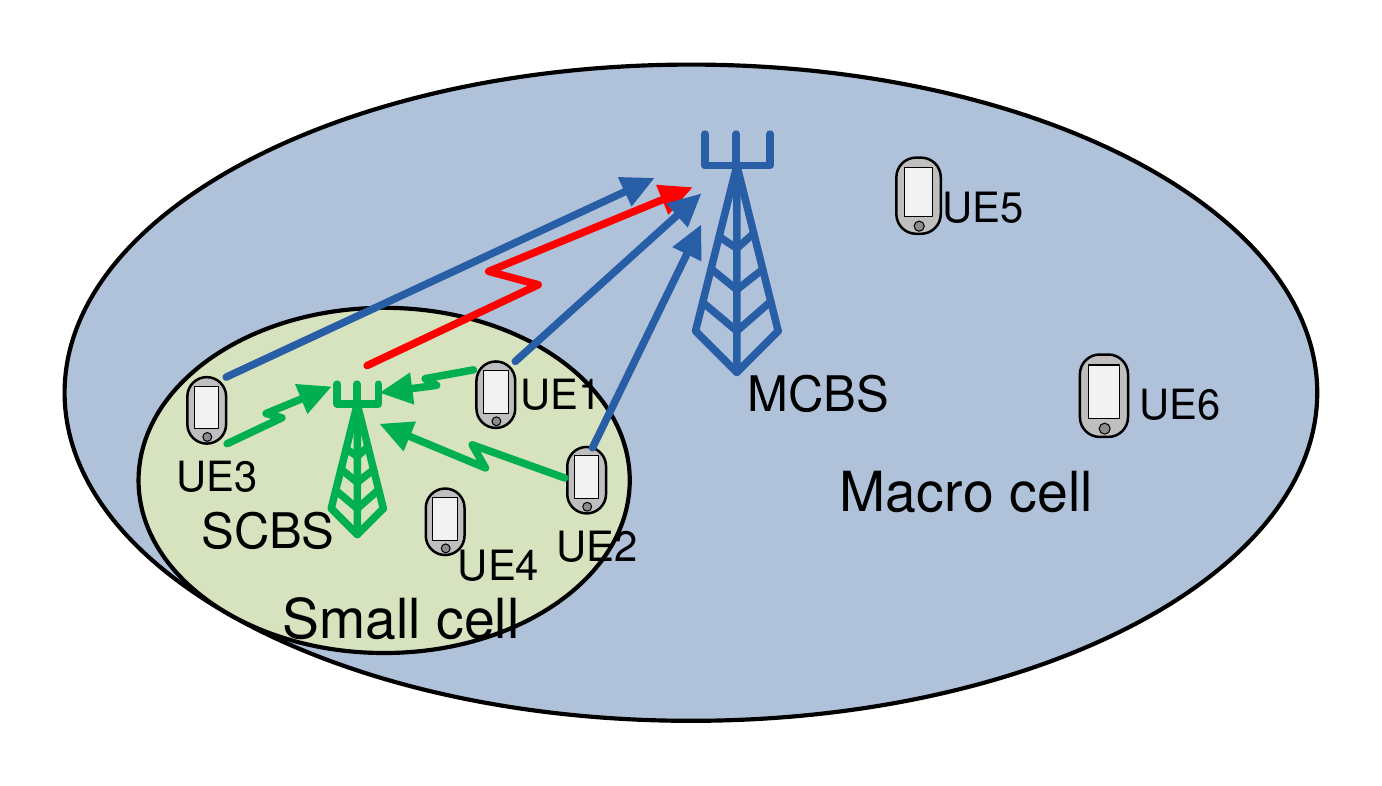}
    \caption{Uplink Transmission in HetNet.} \label{fig:HETNET}
    \end{minipage}
    \vspace*{-2mm}
    \hfill
\begin{minipage}[b]{0.48\linewidth}
    \includegraphics[width=0.75\textwidth]{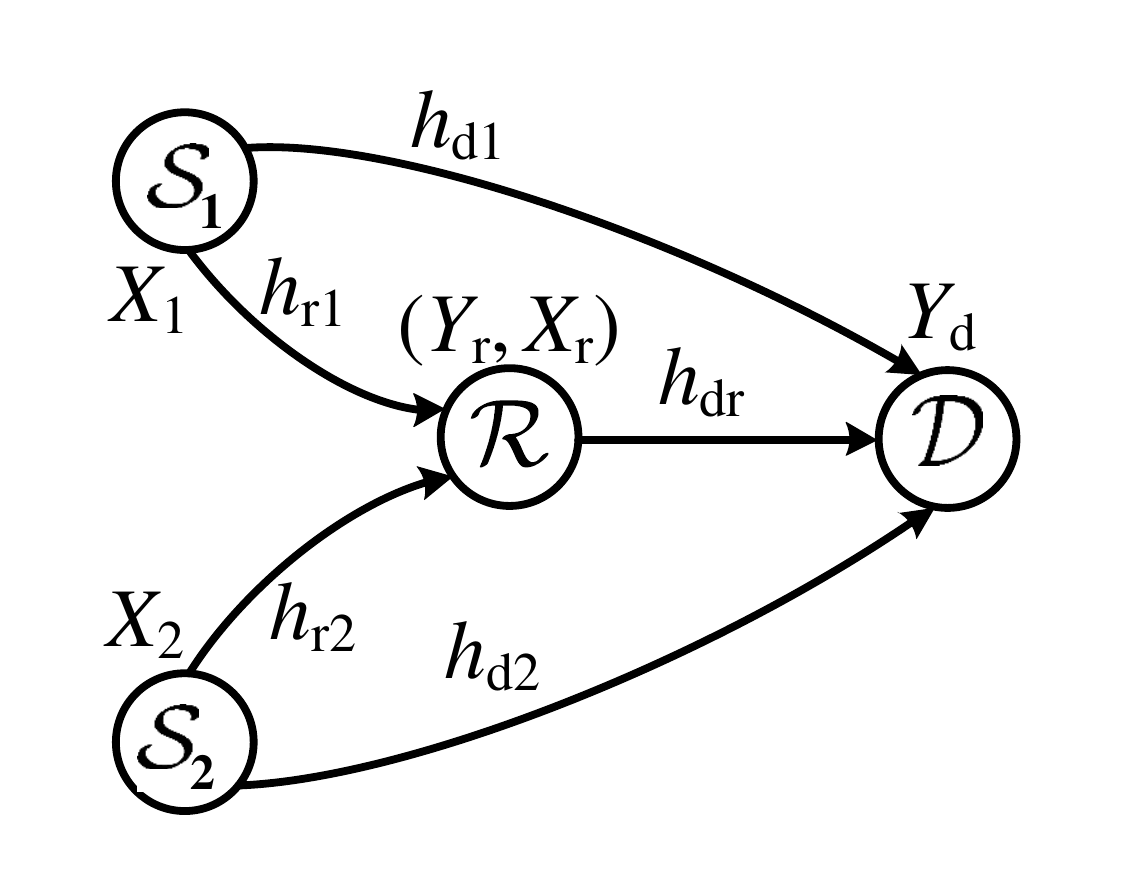}
    \caption{The channel model of full-duplex MARC.} \label{fig:system_model}
    \end{minipage}
    \vspace*{-2mm}
\end{figure}

\textit{$1)$ Small cells:}
 Development of $5$G has been evolving towards ultra-dense networks  (UDN) \cite{BREF} with smaller cell radius and greater cell densification to improve the capacity and coverage of the cellular network \cite{ERK}.
Traditional wired backhaul is no longer cost effective. Instead, wireless backhaul connection between the SCBS and MCBS becomes a more economically viable solution \cite{5GRD}, which has been standardized since LTE A R10 \cite{LTEB}.
Hence, it is important to study the transmission design in HetNets with wireless backhaul connection to maximize the end-to-end performance.

\textit{$2)$ In-band full duplex (FD) transceivers at the SCBS:}  In the current 4G cellular network, the SCBSs relay information between the UEs and the MCBS  in a half-duplex mode \cite{LTEB}, which significantly limits the end-to-end data transmission rate. To potentially double the data rate, in-band FD technology for concurrent transmission and reception has been considered for the 5G network. Despite the self-interference challenge of  FD transmission, several techniques have been developed to mitigate such interference
 including passive suppression and analog and digital cancellation \cite{FDW1, FDW4, FDXWF}. Hence, SCBSs with in-band FD operation \cite{ERK} is expected to be a promising wireless backhaul technology to reach targets for future cellular networks.

\textit{$3)$ Dual-connectivity (DC):} The DC feature in the LTE R12 standard allows a UE to connect to both the SCBS and MCBS \cite{DHLTE13}. DC can improve the HetNet uplink transmissions as the MCBS receives signals from both the SCBS and the UE,  instead of from the SCBS only as in LTE-A \cite{LTEB}, especially when   the UE-MCBS link quality is similar or better than the UE-SCBS link. This may
       occur when the UEs are close to the MCBS, which can happen in an UDN with random locations of the small cells   (not only on the macro-cell boundary) \cite{BREF, ERK}, the distributed antenna deployment for the MCBS \cite{jing}, or a higher MIMO gain at the MCBS than the SCBS.

 \textit{$4)$ QF relaying:} Although decode-and-forward (DF) relaying is used in a HetNet for the current LTE system  \cite{LTEB},  when the UE-MCBS link is similar or stronger than the UE-SCBS link, the QF relaying strategy outperforms the DF relaying strategy \cite{hsce612}. Moreover, massive MIMO can simplify the optimal quantization design at the SCBS as the optimization only depends on the large scale fading and scales with the number of UEs instead of antennas \cite[Myth $9$]{10mth}.

\textit{$5)$ Massive MIMO systems:} With the BSs being equipped with a large antenna array, the massive MIMO  systems can 1) neglect the small scale fading through channel hardening \cite{CHV}; 2) orthogonalize different user transmission through beamforming and allow concurrent transmission without inter-user interference \cite{lars, jing}; 3) achieve close-to-optimal performance with low complexity linear receivers, e.g, a zero forcing (ZF) receiver \cite{lars}; and 4) scale the resource allocation parameters with the number of UEs instead of antennas as in regular MIMO systems \cite{10mth}. Moreover, since massive MIMO arrays can be made rather compact \cite{10mth},\cite{MMS}, they can be implemented at both the MCBS and SCBS.

For the considered HetNet with regular MIMO systems, optimal relaying at the SCBS  is challenging and complicated to achieve. The SCBS needs to know the full CSI from the UEs to the MCBS and  perform joint processing for its received signal from all UEs, as separate processing suffers from inter-user interference.
%
Using the above mentioned massive MIMO properties,
we investigate how massive MIMO can simplify HetNet uplink transmission with QF relaying, and develop a low-complexity, efficient scheme to maximize  rate
performance.
\vspace*{-4mm}
\subsection{Contributions}
We consider uplink transmission in a massive MIMO HetNet where
the SCBS deploys ZF detection  and QF relaying and the MCBS deploys ZF detection and sliding window decoding.
We extend our initial results in \cite{HDLCP1, HDLCP3} and provide the following contributions:

\textit{$1)$ Optimal quantization for the QF scheme with joint decoding (JD)}:  
      We first consider the case of two UEs in the small cell with QF relaying at the SCBS and JD at the MCBS for both UE messages, which we term as the QF-JD scheme. To maximize the rate region, we derived the optimal quantization
        parameters whose size is shown to be equal to the number of UEs instead of antennas as in a traditional MIMO systems. We obtain the quantization parameter values as explicit functions of the large-scale fading \cite{10mth}.

\textit{$2)$ QF with Wyner-Ziv (WZ) binning and time-division (TD) transmission scheme (QF-WZTD)}: We propose a new QF-WZTD scheme that simplifies the QF-JD scheme  by deploying not only WZ binning at the SCBS  \cite{hsce612} but also TD  transmission for each bin index of a UE's quantized data stream. While TD transmission allows separate transmission and  decoding for each bin index, WZ binning facilitates sequential decoding for the bin index, quantization index, and each UE message. To the best of our knowledge, our scheme is the first work that combines QF relaying with WZ binning and TD transmission.

\textit{$3)$ Efficiency of the QF-WZTD scheme}: By comparing  the two schemes, we prove that the QF-WZTD scheme is more efficient, in the sense that it achieves the same rate region as the QF-JD scheme with a much lower complexity. Since the rate region of separate and sequential decoding is bounded by that of joint decoding \cite{hsce612}, we show that the rate region  of the QF-WZTD scheme is maximized by reaching that of the QF-JD scheme with the same optimal quantization noise variances. Furthermore, despite the
extra parameters to be optimized for the QF-WZTD scheme including: the phase durations, power allocation, and  transmission rates for the binning indices, we show that their optimal values are conveniently obtained as direct functions of the optimal quantization noise variances and hence these extra parameters do not complicate the QF-WZTD scheme.
%

 We analyze the complexity in terms of the codebook size and decoding search space at the MCBS. We show that the complexities increase linearly with the number of UEs in the QF-WZTD scheme and exponentially in the QF-JD scheme.

\textit{$4)$ Generalization to $K$ UEs per small cell and a new water-filling algorithm}: We further generalize  our  results for the two-UE per small cell case to a more practical case of $K-$UE  per small cell. The $K$-UE transmission scheme is similar to the two-UE scheme, but the quantization optimization problem is more challenging. We show that this problem can be formulated as a geometric programming (GP) which can be transformed to a convex problem \cite{BYDC}.

\textit{$5)$ A new water-filling algorithm}: For the convex formulation of the optimal quantization problem for  the $K$-UE case, applying the KKT conditions leads to a water-filling type of solution.  We derive the water-level for the optimal quantization  with help of the Descartes' rule of signs \cite{DCRS}, which is different from the conventional water-filling solution \cite{hsce612}.  This water-level of quantization determines  which UE data streams need to be quantized at the SCBS and at what degree (fine or course quantization). Results show that the SCBS performs finer quantization for a UE data stream as the UE-SCBS link becomes stronger than the UE-MCBS link.

In summery, our analysis show how massive MIMO systems simplify the design of QF relaying in HetNets, where a linear coding complexity that grows with  the number of UEs can be achieved
by deploying TD transmission in addition to WZ binning at the SCBS (QF-WZTD scheme). In addition, the optimal quantization parameters and phase durations are obtained in closed forms that only depend on large scale fading. These simplifications, however, does not degrade the rate performance achieved with the schemes of exponential complexity.
\vspace*{-3mm}
\subsection{Related Work}
Massive MIMO and HetNets, as two key enabling technologies for 5G networks, have received significant research interest \cite{MHT1,MHT2,MHT3,MHT4,MHT8,MHT10,MHT9,MHT5,MHT7,MHT6}. For the two-tier HetNet, several interference management \cite{MHT1,MHT2,MHT3,MHT4,MHT8,MHT9,MHT10} and user association \cite{MHT9,MHT5,MHT7} techniques have been proposed to maximize the network throughput \cite{MHT1,MHT2,MHT3,MHT9,MHT5,MHT7} or energy efficiency \cite{MHT4,MHT8,MHT10}. However, these works consider each UE to be associated with one BS (either MCBS or SCBS), while each SCBS performs either independent uplink/dowlink transmission \cite{MHT2,MHT5,MHT8,MHT7} or DF relaying \cite{MHT1,MHT10,MHT9} with equal resource partitioning to its UEs. We, on the other hand, consider the QF relaying scheme for uplink transmission where the MCBS receives signals from both the SCBS and the UEs.
  Using a theoretical MARC model, we investigate how massive MIMO affects the relaying design in the HetNet.

For the MARC, several relaying schemes were proposed based on DF relaying \cite{MARC2,PMARC,HAIJA_RCMRDC2, IETPDF, KRCF_MARC, HDLCP2} and  QF relaying \cite{KRCF_MARC, CFMARC2, QFMARC, QFMARC2, QFMARC3, CFMARC10, NNC}. QF relaying has different forms, depending on whether the relay deploys WZ binning \cite{hsce612} which partitions the quantization indices into equal size groups with a different bin index for each group.
On one hand, without WZ binning\footnote{The QF scheme with WZ binning is also know as a compress-forward scheme\cite{hsce612}.}, the relay transmits the quantization indices, while the destination jointly  decodes the transmitted messages and quantization indices  by simultaneously using the received signals either over two consecutive blocks \cite{PVUWZ} or whole transmission blocks, as in the noisy network coding (NNC) scheme\cite{NNC}. On the other hand, with WZ binning, the relay transmits the binning indices, while the destination  sequentially  decodes the binning, quantization indices and then the transmitted messages by successfully using the received signals over two consecutive blocks \cite{hsce612}.
While sequential decoding is simpler than joint decoding, it generally leads to a smaller rate region  \cite{NNC}. For the MARC and cloud radio access network (C-RAN)\footnote{C-RAN can be modeled as MARC with multiple relays but without the direct links from the users to the destination}\cite{LEIS2, WEIYU2}, sequential decoding leads to the same sum rate as the joint decoding but a different rate region  \cite{LEIS2, WEIYU2}. However, in this paper, besides WZ binning and sequential decoding, we utilize massive MIMO properties to further simplify the QF relaying scheme by deploying TD transmission from the SCBS to the MCBS, which 1) reduces the codebook size of the transmitted signals by the SCBS, 2) allows separate and sequential decoding at the MCBS, and 3) at the same time, achieves the same rate region of the more complex joint decoding schemes, instead of only the sum rate as in \cite{LEIS2}.

Besides developing transmission and decoding techniques of QF relaying schemes, the quantization resolution at the relay is a key design problem for rate maximization. For multi-antenna QF relaying,   it is  equivalent to
the quantization noise covariance optimization problem, which is
in general non-convex and challenging to solve. Hence, approximate solutions have been obtained via iterative numerical methods for one way \cite{cmv1} and two way \cite{cmv2}  half-duplex relay channels and C-RAN \cite{WEIYU,WEIYU2}.
With massive MIMO considered in this paper, the quantization problem can be simplified with possible closed-form solutions, avoiding high computational complexity and processing delay. Moreover, without any degradation of the rate performance, massive MIMO facilitates the TD transmission  for different information from the SCBS to the MCBS.
%
%
%
\vspace*{-3mm}
\subsection{Organization}
The remainder of this paper is organized as follows. Section \ref{sec:system_model} presents the uplink channel model of a massive MIMO HetNet. 
Section \ref{sec: simpsch} presents the  QF-JD transmission scheme for two UEs  and provides its achievable rate region. To maximize this region, Section \ref{OPQF} derives the optimal quantization at the SCBS.
Section \ref{sec: sss}  presents the low-complexity QF-WZTD scheme  and derives its achievable rate region. Section \ref{comps2}   provides the comparison of the two schemes in complexity and achievable rate region.
Section \ref{sec: gen} generalizes the two schemes to the
$K$-UE scenario. Section \ref{sec:cap. gau} presents numerical results, and Section \ref{sec:conclusion} concludes the paper.
\vspace*{-4mm}
\section{System Model}\label{sec:system_model}
We consider the uplink transmission in a HetNet that consists of a macro cell, a small cell, and $K$ UEs $(K>1)$ in the small cell.
Each UE has a single antenna while
the SCBS and MCBS have $N$ and $M$ antennas, respectively, where we assume $M\gg N\gg K$. Since using massive MIMIO techniques at the MCBS and SCBS can reduce the uplink interference from other nodes to a negligible level \cite{lars}, we ignore transmissions in and from other SCBSs in the same macro cell.  For simplicity, we start with two UEs (UE1 and UE2) and then generalize the results to $K>2$ UEs in Section \ref{sec: gen}. The two UEs communicate with the MCBS with the help of the  SCBS, as shown in Fig. \ref{fig:HETNET}. This uplink channel resembles the  MARC shown in Fig. \ref{fig:system_model}, where the SCBS  resembles the relay (denoted by ${\cal R}$) and the MCBS resembles  the destination (denoted by ${\cal D}$).

For  the MARC in Fig. \ref{fig:system_model}, we assume a block fading channel model where the channel over each link remains constant in each transmission block and changes independently between blocks. Over  $B$ transmission blocks where $B\gg 1$, let $\mathbf{h}_{rk,j}=[h_{rk,j}^{(1)},\cdots,h_{rk,j}^{(N)}]^T$ denote the $N\times 1$  channel vector from $\text{UE}_k$ to ${\cal R}$  in block $j,$ for $k\in\{1,2\}$ and $j\in\{1,\ldots,B\}$, where $h_{rk,j}^{(n)}$ is the channel coefficient from $\text{UE}_k$ to the $n^{\text{th}}$ antenna of ${\cal R}$ in block $j$.  We assume $\mathbf{h}_{rk,j}$ is a complex Gaussian random vector with zero mean
and covariance $\sigma_{h,r}^2 {\bf I}$. The variance  $\sigma_{h,r}^2 = d_{rk}^{\alpha}$  is based on the pathloss model, where $d_{rk}$ is the distance between $\text{UE}_k$ and ${\cal R}$, and $\alpha$ is the pathloss exponent.
A Similar definition holds for the $M\times 1$ channel vector $\mathbf{h}_{di,j}$ from $\text{UE}_k$ to ${\cal D}$ and the $M\times N$ channel matrix $\mathbf{H}_{dr}$ from ${\cal R}$ to ${\cal D}$.
We assume all channel coefficients are independent of each other.

At any transmission block $j\in\{1,\ldots,B\}$, given $\mathbf{h}_{rk,j}$, $\mathbf{h}_{dk,j},$ and $\mathbf{H}_{dr,j}$, the received signal vectors at ${\cal R}$ and ${\cal D}$, denoted by $\mathbf{y}_{r,j}$ and $\mathbf{y}_{d,j}$ respectively, are
given as follows:
\noindent
\begin{align}\label{gaumod}
\mathbf{y}_{r,j}&=\mathbf{h}_{r1,j}x_{1,j}+\mathbf{h}_{r2,j}x_{2,j}+\mathbf{z}_{r,j},\quad
\mathbf{y}_{d,j}=\mathbf{h}_{d1,j}x_{1,j}+\mathbf{h}_{d2,j}x_{2,j}+\mathbf{H}_{dr,j}\mathbf{x}_{r,j}+\mathbf{z}_{d,j},
\end{align}
\noindent where  $x_{k,j}$ is the transmit signal by $\text{UE}_k$ for $k\in\{1,2\}$ while $\mathbf{x}_{r,j}$ is the $N\times 1$ transmit signal vector from ${\cal R}$;
$\mathbf{z}_{r,j}$ and $\mathbf{z}_{d,j}$
are $N\times1$ and $M\times1$ independent complex AWGN vectors with zero mean and covariance ${\bf I}_N$ and ${\bf I}_M$, respectively.


We assume that the channel state information (CSI) is known at the respective receivers, i.e.,  ${\cal R}$ knows $\mathbf{h}_{rk}$ and ${\cal D}$ knows $\mathbf{h}_{dk}$ and $\mathbf{H}_{dr}$. Moreover, ${\cal R}$ knows (via feedback from ${\cal D}$ \cite{TBF}) the variance of the channel from ${\cal R}$ to ${\cal D}$, and from each UE to ${\cal D}$, through the pathloss information. Such knowledge helps ${\cal R}$ optimize its transmission for maximum rate region (see Sections \ref{OPQF} and \ref{sec: sss}).
Note that the pathloss information over each link is much easier to obtain than the massive MIMO channel itself
($\mathbf{h}_{dk}$ and $\mathbf{H}_{dr}$) in each block.

We consider full-duplex relaying at the SCBS. Although full-duplex relaying suffers from  self-interference,  it can be substantially alleviated by analog and digital cancellation techniques  developed in recent research while the
remaining part appears as additional additive
noise \cite{FDXWF}. Hence, we ignore this interference  and focus on its transmission design.
 %

In massive MIMO systems, linear detectors like   ZF or maximum ratio
combining (MRC) are applied to reduce the inter-user interference  \cite{lars}. In this paper, we choose the ZF detector for simplicity. However, similar analysis is applicable to other detectors. Receivers at ${\cal R}$ and ${\cal D}$ apply ZF  detection to separate the data streams from different origins.   Specifically,
define the ZF matrices $\mathbf{A}_{r,j}$ and $\mathbf{A}_{d,j}$ as follows:
\begin{align}\label{zfff}
\mathbf{A}_{r,j}&\triangleq(\mathbf{G}_{r,j}^{H}\mathbf{G}_{r,j})^{-1}\mathbf{G}_{r,j}^{H},
\;\mathbf{A}_{d,j}\triangleq(\mathbf{G}_{d,j}^{H}\mathbf{G}_{d,j})^{-1}\mathbf{G}_{d,j}^{H},\nonumber\\
\text{where}\;
\mathbf{G}_{r,j}&\triangleq[\mathbf{h}_{r1,j}\;\mathbf{h}_{r2,j}],\;
\mathbf{G}_{d,j}\triangleq[\mathbf{h}_{d1,j}\;\mathbf{h}_{d2,j}\;\mathbf{H}_{dr,j}].
\end{align}
%
%
  Denote $\mathbf{A}_{r,j}=[\mathbf{a}_{r1,j},\mathbf{a}_{r2,j}]^H$ and $\mathbf{A}_{d,j}=[\mathbf{a}_{d1,j},\mathbf{a}_{d2,j}, \mathbf{A}_{dr,j}]^H$, where $\mathbf{a}_{rk,j}$ is an $N\times 1$ vector,  $\mathbf{a}_{dk,j}$ is an $M\times 1$ vector, for $k\in\{1,2\}$, and $\mathbf{A}_{dr,j}$ is an $M\times N$ matrix. After applying ZF matrices in (\ref{zfff})
 to the received signal vectors $\mathbf{y}_{r,j}$ and $\mathbf{y}_{d,j}$  at ${\cal R}$ and ${\cal D}$ in (\ref{gaumod}), we obtain the $2\times 1$ received vector $\tilde{\mathbf{y}}_{r,j}$ at ${\cal R}$ and $(2+N)\times 1$ received vector $\tilde{\mathbf{y}}_{d,j}$ at ${\cal D}$ as follows:
\begin{align}\label{gaumodzf}
\tilde{\mathbf{y}}_{r,j}&=\left[\tilde{y}_{r1,j}\; \tilde{y}_{r2,j}\right]^T,\;\tilde{\mathbf{y}}_{d,j}=\left[\tilde{y}_{d1,j}\; \tilde{y}_{d2,j}\; \tilde{\mathbf{y}}_{dr,j}\right]^T, \nonumber\\
\tilde{y}_{rk,j}&=x_{k,j}+\mathbf{a}_{rk,j}^{H}\mathbf{z}_{r,j},
\;\tilde{y}_{dk,j}=x_{k,j}+\mathbf{a}_{dk,j}^{H}\mathbf{z}_{d,j},\;k\in\{1,2\},\;
\tilde{\mathbf{y}}_{dr,j}=\mathbf{x}_{r,j}+\mathbf{A}_{dr,j}^{H}\mathbf{z}_{d,j},
\end{align}
where $\tilde{y}_{rk,j}$ and $\sim\tilde{y}_{dk,j}$ are the signals received at ${\cal R}$ and$\sim{\cal D},$ respectively, from $\text{UE}_k$ in block $j$, and $\tilde{\mathbf{y}}_{dr,j}$ is the $N\times1$ signal vector received at ${\cal D}$ from ${\cal R}$ in block $j$.
By (\ref{gaumodzf}), we see that the massive MIMO system with ZF detection transforms the general MARC in (\ref{gaumod}) to a MARC with orthogonal receivers at ${\cal R}$ and ${\cal D}$ instead of ${\cal D}$ only as in \cite{hsce612, WEIYU, CFMARC2}.
\vspace*{-3mm}
\section{The QF-JD Scheme for Massive MIMO HetNet}\label{sec: simpsch}
This transmission scheme is based on QF relaying at the SCBS and JD at the MCBS, which we term the QF-JD scheme. Since massive MIMO asymptotically orthogonalizes the transmission from different users \cite{jing, lars}, the SCBS can   quantize separately the data stream from each UE as in \cite{hsce612}. The MCBS then utilizes the signals received from both UEs and the SCBS  to decode both UE messages.
The QF-JD scheme is designed by modifying the QF schemes in \cite{hsce612, QFMARC} to the HetNet channel model in (\ref{gaumodzf}). The data transmission from UEs to the MCBS is carried over $B$ transmission blocks where each UE sends $B$$-$$1$ messages through these blocks.\footnote{ This may reduce the average rate region  by a factor of (1/B), but this factor becomes negligible as $B\rightarrow \infty$ \cite{hsce612}.}
In each block $j$, each UE transmits a new message; the SCBS quantizes the  data stream received from each UE in (\ref{gaumodzf}), and sends the quantization indices to the MCBS in the next block $j+1$, then the MCBS applies  sliding window decoding over two consecutive blocks.  We describe the transmission scheme in more detail below. 
\vspace*{-4mm}
\subsection{The QF-JD  Transmission Scheme}\label{secsigsc}
 In  transmission block $j\in\{1,2,...,B\}$, $\text{UE}_1$ sends its new message $w_{1,j}$ by transmitting its codeword $U_1(w_{1,j})$. Similarly, $\text{UE}_2$ sends $w_{2,j}$ by transmitting $U_2(w_{2,j})$. At the end of block $j$, the SCBS first deploys ZF detection for its received signal $\mathbf{y}_{r,j}$. Then, it quantizes the detector's output signals $\tilde{y}_{r1,j}$ and $\tilde{y}_{r2,j}$ and determines the quantization indices $(l_{1,j},l_{2,j})$. Finally, the SCBS generates a common codeword $\mathbf{U}_r(l_{1,j},l_{2,j})$ for both indices $(l_{1,j},l_{2,j})$ and transmits it in block $j+1$ to the MCBS. 
\subsubsection{Transmit Signals}
During transmission block $j$, $\text{UE}_1$ (resp. $\text{UE}_2$) transmits the signal $x_1(w_{1,j})$ (resp. $x_2(w_{2,j})$), and the SCBS  transmits the signal $\mathbf{x}_r(l_{1,j-1},l_{2,j-1})$ (for signals received from previous block $j-1$). These signals are constructed as follows:
\begin{align}\label{sigG}
x_{1,j}=&\;\sqrt{P_1}U_1(w_{1,j}),\;x_{2,j}=\sqrt{P_2}U_2(w_{2,j}),\;
\mathbf{x}_{r,j}=\sqrt{P_r/N}\mathbf{U}_r(l_{1,j-1},l_{2,j-1}),
\end{align}
 where $U_1(w_{1,j})$ and $U_2(w_{2,j})$  are i.i.d Gaussian signals with zero mean and unit variance and they respectively convey the codewords of the UE messages $w_{1,j}$ and $w_{2,j}$, and
 $\mathbf{U}_r(l_{1,j-1},l_{2,j-1})$ is an $N\times 1$ Gaussian random vector with zero mean and covariance ${\bf I}_N$, which conveys the codeword of the quantization index pair $(l_{1,j-1},l_{2,j-1})$. The transmit powers at $\text{UE}_1$, $\text{UE}_2$,  and the SCBS are $P_1,$ $P_2$ and $P_r$, respectively.
  Moreover, after ZF processing at the SCBS, we obtain the signals in (\ref{gaumodzf}). After quantization, we have:
 \begin{align}\label{sigQFG}
\hat{y}_{r1,j}=&\;\tilde{y}_{r1,j}+\hat{z}_{r1,j},\;\hat{y}_{r2,j}=\tilde{y}_{r2,j}+\hat{z}_{r2,j},
\end{align}
 where $\hat{y}_{rk,j},$ for $k\in\{1,2\}$ is  the quantized version of $\tilde{y}_{rk,j}$ in (\ref{gaumodzf}) and $\hat{z}_{rk,j}$ is the quantization noise with zero mean and $Q_k$ variance $\sim\mathcal{CN}(0,Q_k)$.
\subsubsection{Decoding}
The MCBS decodes the messages of both UEs using joint typicality (JT) \cite{hsce612} or maximum likelihood (ML) \cite{abramowitz1972hmf} decoding methods. Here, we briefly describe the decoding techniques
 while the theoretical analysis is given in subsection 3 of Appendix A.

Specifically, the MCBS performs sliding window decoding over two  consecutive blocks ($j$ and $j+1$) to decode both UEs' messages. At the end of block $j+1$, after performing ZF detection in (\ref{gaumodzf}), the MCBS simultaneously utilizes the received signals directly from the UEs in block $j$ ($\tilde{y}_{d1,j}$ and $\tilde{y}_{d2,j}$) and from the SCBS in block $j+1$ ($\tilde{\mathbf{y}}_{dr,j+1}$) to jointly decode both UE  messages $(w_{1,j},w_{2,j})$ for some quantization indices $(l_{1,j},l_{2,j})$.
\vspace*{-4mm}
\subsection{Achievable Rate Region}\label{sec:achrr}
Let $R_1$ and $R_2$ denote the transmission rates for $\text{UE}_1$ and $\text{UE}_2$, respectively. The achievable rate region  is determined by the rate constraints that ensure reliable decoding at the MCBS. These constraints are derived from the error analysis of the decoding rule at the MCSB as follows.
\begin{thm}\label{thss}
For the considered massive MIMO HetNet, the rate region for $\text{UE}_1$ and $\text{UE}_1$ achieved by the QF-JD scheme consists of all rate pairs $(R_1, R_2)$ satisfying:
\begin{align}\label{dmcac}
R_1&\leq \min\{I_1,I_2\},\; R_2\leq \min\{I_3,I_4\},\;
R_1+R_2\leq I_5,
\end{align}
where
\vspace{-3mm}
\begin{align}\label{sth1rr1}
I_1=&\;{\cal C}\left(\frac{P_1(M-N)}{d_{d1}^{\alpha}}+\frac{P_1}{(d_{r1}^{\alpha}/N)+Q_1}\right),\;
I_2={\cal C}\left(\frac{P_1(M-N)}{d_{d1}^{\alpha}}\right)+\zeta,\\
I_3=&\;{\cal C}\left(\frac{P_2(M-N)}{d_{d2}^{\alpha}}+\frac{P_2}{(d_{r2}^{\alpha}/N)+Q_2}\right),\;I_4={\cal C}\left(\frac{P_2(M-N)}{d_{d2}^{\alpha}}\right)+\zeta,\nonumber\\
I_5=&\;{\cal C}\left(\frac{P_1(M-N)}{d_{d1}^{\alpha}}\right)+{\cal C}\left(\frac{P_2(M-N)}{d_{d2}^{\alpha}}\right)+\zeta,\;
\zeta= N{\cal C}\left(\frac{P_r(M-N)}{Nd_{dr}^{\alpha}}\right)
-{\cal C}\left(\frac{d_{r1}^{\alpha}}{NQ_1}\right)-{\cal C}\left(\frac{d_{r2}^{\alpha}}{NQ_2}\right),\nonumber
\end{align}
 where ${\cal C}(x)\triangleq \log(1+x)$. The transmission rates ($R_{q1}$ and $R_{q2}$) for the quantization indices $(l_1,l_2)$ are given by:
\begin{align}\label{qrt1}
R_{qk}&={\cal C}\left(\frac{1+P_k\big(N/d_{rk}^{\alpha}\big)}{Q_kN/d_{rk}^{\alpha}}\right),\quad k\in\{1,2\}.
\end{align}
\end{thm}
\IEEEproof
The constraints in (\ref{dmcac}) ensure reliable decoding at the MCBS.  Specifically, $I_1$ (resp. $I_2$) ensures reliable decoding of $\text{UE}_1$ message (resp. $\text{UE}_1$ message and the quantization indices) while receiving $\text{UE}_2$ message and the quantization indices (resp. $\text{UE}_2$ message) correctly.  Similar explanation holds for $I_3$ and $I_4$.
 Finally, $I_5$ ensures reliable decoding of both messages and quantization indices.
 (A detailed proof is given in Appendix A.)
\endIEEEproof
\begin{rem}
The same rate region given in Theorem \ref{thss} is achieved by the NNC scheme \cite{NNC} by applying the results of Theorem 1 in \cite{NNC} to the HetNet channel model in (\ref{gaumodzf}). Instead of jointly decoding all the transmitted messages by simultaneously using the received signals either over two consecutive blocks as in the QF-JD scheme or whole transmission blocks as in the NNC scheme \cite{NNC}, the proposed QF-WZTD scheme in Section \ref{sec: sss}  utilizes WZ binning and  TD transmission from the SCBS to the MCBS to reduce the codebook size of the transmitted signals by the SCBS, and allow separate and sequential decoding at the MCBS without reducing the achievable rate region.
\end{rem}
\begin{rem}
The results in Theorem \ref{thss} show that the massive MIMO system simplifies the quantization process at the SCBS as compared with a regular MIMO system. Unlike a regular MIMO system which requires optimizing the covariance matrix of the quantization noise vector \cite{cmv1,cmv2} to obtain the rate region boundaries, the massive MIMO system only requires two quantization elements ($Q_1$ and $Q_2$) to be optimized.
 This simplification coincides with \cite[Myth $9$]{10mth} that in the massive MIMO system the complexity of resource allocation scales with the number of UEs instead of the number of antennas.
 \end{rem}
\vspace*{-3mm}
\section{Optimal Quantization $(Q_1^{\ast},Q_2^{\ast})$  at the SCBS}\label{OPQF}
 It is important to specify the optimal quantization  at the SCBS for each UE data stream. As the quantization levels increase, the quantizer becomes finer with smaller noise variances $(Q_1,Q_2)$ at its outputs $(\hat{y}_{r1,j},\hat{y}_{r2,j})$ in (\ref{sigG}). However, finer quantization requires a higher transmission rate for the quantization indices, which may not be sustained by the link from the SCBS to the MCBS.  In this section, we derive quantization parameters $(Q_1^{\ast},Q_2^{\ast})$ that optimize the rate region in (\ref{dmcac}).

In Theorem \ref{thss}, any boundary point of the rate region can be represented by the weighted sum rate $\mu_1 R_1 +\mu_2 R_2$, where $\mu_1\in [0,1]$ is some priority weighting factor of the $\text{UE}_1$ rate, and $\mu_2 = 1-\mu_1$.  Thus, the rate region boundary is achieved by maximizing the weighted sum rate for some given $\mu_1$ over $Q_1$ and $Q_2$. Hence, the optimization problem is formulated as:
%
\begin{align}\label{WSRO}
\max_{\substack{Q_1, Q_2}} &\mu_1 R_1 +\mu_2 R_2,\\
\text{s.t.}\; & R_1\leq \min\{I_1,I_2\},\; R_2\leq \min\{I_3,I_4\},\;R_1+R_2\leq I_5,\;
Q_1\geq0,\;Q_2\geq0,\nonumber
\end{align}
where $I_1,I_2,\ldots,I_5$ are given in (\ref{sth1rr1}) with full transmission powers in (\ref{sigG}). The solution of problem (\ref{WSRO}) is given as follows:
\begin{thm}\label{cor1}
The optimal $Q_1^{\ast}$ and $Q_2^{\ast}$ for  problem (\ref{WSRO}) are given by:
\begin{align}\label{opsI1}
Q_k^{\ast}&=\frac{(d_{rk}^{\alpha}/N)\big(1+\frac{P_k(M-N)}{d_{dk}^{\alpha}}\big)+P_k}
{\big(1+\frac{P_k(M-N)}{d_{dk}^{\alpha}}\big)(\lambda_k^{\ast}-1)},\;k\in\{1,2\},
\end{align}
\begin{align}
\text{where}\;\lambda_1^{\ast}&=\min\left\{\max\{\lambda_o,1\},\lambda_s\right\},\quad \lambda_2^{\ast}=\lambda_s/\lambda_1^{\ast},\;
\lambda_o=(2A)^{-1}(B+\sqrt{B^2-4AC}),\quad ,\nonumber\\
A&=\mu_2P_2(d_{r1}^{\alpha}/N)\big(1+\frac{P_1(M-N)}{d_{d1}^{\alpha}}\big),\;
C=-\mu_1 P_1(d_{r2}^{\alpha}/N)\big(1+\frac{P_2(M-N)}{d_{d2}^{\alpha}}\big)\lambda_s,\nonumber\\
B&=(\mu_1-\mu_2)P_1P_2,\; \text{and}\; \lambda_s=\left(1+\frac{P_r(M-N)}{Nd_{dr}^{\alpha}}\right)^N.\nonumber
\end{align}
If $\lambda_o<1$ (resp. $\lambda_o>\lambda_s)$, set $\lambda_1^{\ast}=1$ (resp. $\lambda_1^{\ast}=\lambda_s)$.
\end{thm}
\IEEEproof
We prove the result in the following steps:
\begin{enumerate}[leftmargin=*]
  \item For $\mu_1\in (0.5,1]$, the weighted sum rate $R_{ws}=\mu_1 R_1 +\mu_2R_2$ can be expressed as follows:
      \begin{align}\label{WSM}
        \!\!\!\!&R_{ws}=(\mu_1-\mu_2)R_1+\mu_2(R_1+R_2)\stackrel{a}{\leq} (\mu_1-\mu_2)\min\{I_1,I_2\}+\mu_2 \min\{I_1+I_3,I_5\},\\
        \!\!\!\!&\stackrel{b}{=}\left\{\begin{array}{cl}
             \!\!\!\!\mu_1 I_1 +\mu_2 I_3 & \!\!\text{if}\; I_1\leq I_2,\;\&\; I_1+I_3\leq I_5, \\
             \!\!\!\!(\mu_1-\mu_2)I_2+\mu_2 I_5 & \!\!\text{if}\; I_1> I_2,\;\&\; I_1+I_3> I_5, \\
             \!\!\!\!(\mu_1-\mu_2)I_2+\mu_2 (I_1+I_3) &\!\!\text{if}\;I_1> I_2,\;\&\; I_1+I_3\leq I_5,\\
             \!\!\!\!(\mu_1-\mu_2)I_1+\mu_2 I_5& \!\!\text{if}\;I_1\leq I_2,\;\&\; I_1+I_3> I_5
            \end{array}\right.,\nonumber
      \end{align}
      where $(a)$ follows from the rate constraints in (\ref{WSRO}) while $(b)$ follows from all four possible cases of $(a)$ at any $Q_1$ and $Q_2$ and by noticing that $I_1+I_4>I_5$ and $I_2+I_3>I_5$.
  \item We maximize each case in (\ref{WSM}.b) subject to its constraints and then choose the case that maximizes  $R_{ws}$. Considering (\ref{dmcac}), for:
      \begin{itemize}[leftmargin=*]
        \item Case 1: the constraint $I_1\leq I_2$ is redundant when $I_1+I_3\leq I_5$.
        Hence, the optimization problem becomes:
            \begin{align}\label{EWSM}
             \!\!\!\!\!\!\!\!\!\!\!\!\!\!\!\! \max_{\substack{Q_1, Q_2}}   \mu_1 I_1 +\mu_2 I_3, \;\text{s.t.}\;I_1+I_3= I_5,\;Q_1\geq0,\; Q_2\geq 0.
            \end{align}
            Since $I_{1}$ $(I_3)$ is a decreasing function with $Q_1$ $(Q_2)$ while $I_5$ is increasing with both $Q_1$ and $Q_2$, $R_{ws}$ is maximized when $I_1+I_3\leq I_5$ holds with equality.
        \item Case 2 is equivalent to  individual rate maximization of $\text{UE}_1$ (denoted as $R_1^{\max}$), since  $I_1+I_3> I_5$ is redundant when $I_1>I_2$. Moreover, $R_{ws}$ is independent of $Q_2$ since $(\mu_1-\mu_2)I_2+\mu_2 I_5$ $=\mu_1I_2+\mu_2  {\cal C}\left((P_2(M-N))/(d_{d2}^{\alpha})\right)$.
        \item Case 3 is also equivalent to $R_1^{\max}$, since besides the constraints $I_1+I_3\leq I_5$ and $I_1> I_2,$ it is clear from (\ref{dmcac}) that $I_5-I_2\leq I_3$. Hence, these three inequalities only hold when they are equal, i.e. $I_5-I_3=I_2=I_1$, which is possible only when $Q_2\rightarrow \infty$, i.e., $R_1^{\max}$.
        \item In Case 4, $R_{ws}=(\mu_1-\mu_2)I_1+\mu_2 I_5$ is maximized by maximum possible value of $Q_2$, which is obtained when $I_1+I_3= I_5$.
      \end{itemize}
      Therefore, considering the four cases, the optimization problem in (\ref{WSRO}) is equivalent to (\ref{EWSM}).
  \item Considering $I_5$ in (\ref{EWSM}), let
  \begin{align*}
  N{\cal C}\left(\frac{P_r(M-N)}{Nd_{dr}^{\alpha}}\right)=\log(\lambda_s)=\log(\lambda_1)+\log(\lambda_2),
  \end{align*}
  where $\lambda_s$ is given in (\ref{opsI1}) while $\lambda_1 \lambda_2=\lambda_s$. Then, with $I_1+I_3=I_5,$ we get $Q_1$ (resp. $Q_2$) as $Q_1^{\ast}$ (resp. $Q_2^{\ast}$) in (\ref{opsI1}) except that $\lambda_1$ (resp. $\lambda_2$) is not yet optimized. $\lambda_1$ and $\lambda_2$ are optimized in step $5$.
  \item By substituting $Q_1$ and $Q_2$ given in step $4$ into (\ref{EWSM}), the optimization problem in (\ref{EWSM}) becomes as follows:
      \begin{align}\label{EWSMU}
              &\max_{\substack{\lambda_1, \lambda_2}}   \mu_1 \bar{I}_1 +\mu_2 \bar{I}_3,\quad \text{s.t.}\;\lambda_1 \lambda_2=\lambda_s,\;\lambda_1\geq 1,\;\lambda_2\geq 1, \\
       \text{where}\;&\bar{I}_1={\cal C}\left(\frac{P_1(M-N)}{d_{d1}^{\alpha}}\right)+{\cal C}\left(\frac{P_1(M-N)}{d_{d1}^{\alpha}}+\frac{P_1N}{d_{r1}^{\alpha}}\right)-{\cal C}\left(\frac{P_1(M-N)}{d_{d1}^{\alpha}}+\frac{P_1N}{\lambda_1d_{r1}^{\alpha}}\right),\nonumber
      \end{align}
      which is obtained from $I_1$ in (\ref{sth1rr1}) with $Q_1$ given in step $4$, $\bar{I}_3$ has similar expression to $\bar{I}_1$ except for switching each subscript from $1$ to $2$ and the condition $\lambda_k\geq1$ ensures that $Q_k\geq 0$. Since $\bar{I}_1$ depends on $\lambda_1$ only in the negative term,  (\ref{EWSMU}) can be reexpressed as follows:       
      \begin{align}\label{EWSMUU}
             &\min_{\substack{\lambda_1, \lambda_2}}   \sum_{k=1}^{2}\mu_k\log\left(1\!+\!\frac{P_k(M-N)}{d_{dk}^{\alpha}}+\frac{P_kN}{\lambda_kd_{rk}^{\alpha}}\right),
              \;\text{s.t.}\;\lambda_1 \lambda_2=\lambda_s,\;\lambda_1\geq 1,\;\lambda_2\geq 1. 
      \end{align}
      \item By substituting $\lambda_2=\lambda_s/\lambda_1$ into (\ref{EWSMUU}) and then deriving (\ref{EWSMUU}) with respect to $\lambda_1$, we obtain $\lambda_1^{\ast}$ as a solution of 
          \begin{align}\label{funl}
          f_1(\lambda_1)=0, \; \text{where}\;f_1(\lambda_1)=A\lambda_1^2-B\lambda_1+C,
          \end{align}
          while $A,B$ and $C$ are given in (\ref{opsI1}). If $\lambda_1$ from (\ref{funl}) is $<1$ (resp. $>\lambda_s)$, the function in (\ref{EWSMUU}) is increasing (resp. decreasing) over  $\lambda_1\in[1,\lambda_s]$, then $\lambda_1^{\ast}=1$ (resp. $\lambda_s)$.
\end{enumerate}
\endIEEEproof
\begin{rem}\label{fineQ}
Theorem \ref{cor1} has several implications:
 \begin{itemize}[leftmargin=*]
   \item The rate region boundary:
   \begin{itemize}[leftmargin=*]
      \item Maximum individual rate $R_1^{\max}$ (resp. $R_2^{\max}$) is achieved by setting $(\mu_1,\mu_2)=(1,0)$ (resp. $(\mu_1,\mu_2)=(0,1)$). With $\mu_2=0$, $f_1(\lambda_1)$ in (\ref{funl}) is decreasing with $\lambda_1$. Hence, $(\lambda_1^{\ast},\lambda_2^{\ast})=(\lambda_s,1)$. Consequently, $Q_1^{\ast}$ is as in (\ref{opsI1}) and $Q_2^{\ast}= \infty$ (i.e., not relaying the signal from $\text{UE}_2$). Therefore, $R_1^{\max}$ is achieved when the SCBS  optimally quantize and forward the received signal from $\text{UE}_1$, but ignores the signal from $\text{UE}_2$.
           In this case, the $\text{UE}_2$ signal is received at the MCBS through the direct link only. 
     \item Maximum sum rate is achieved  by setting $\mu_1=\mu_2=0.5$.
   \end{itemize}
   \item The SCBS performs very fine quantization for at least one UE signal since both $Q_1^{\ast},$ and $Q_2^{\ast}$ are inversely proportional to $(1+\text{SNR}_{rd})^{0.5N}$, where $N\gg 1$ and $\text{SNR}_{rd}$ is the SNR from the SCBS to the MCBS.
   \item The optimal quantization depends on the large-scale fading (e.g., pathloss) of each link. The MCBS can feedback the pathloss information to the SCBS, which is much easier to obtain than the instantaneous CSI of  all channel vectors. Hence, for the massive MIMO HetNet, the transmission of none-UE data can be minimized, which satisfies the ultra-lean design requirement of 5G networks \cite{ERK}.
 \end{itemize}
\end{rem}
\vspace*{-4mm}
\section{The QF-WZTD Scheme}\label{sec: sss}
The QF-JD scheme in Section \ref{sec: simpsch} deploys 1) a common codeword $\mathbf{U}_r$ for both quantization indices at the SCBS to be transmitted to the MCBS, and 2) JD of both UE messages at the MCBS. The resulting  computational complexity is exponential with respect to the number of UEs at both the SCBS and  MCBS \cite{hsce612} (see Section \ref{sec: COMP} for details).
Briefly speaking, let $L_k$ be the set of quantization indices for the $\text{UE}_k$ data stream where $k\in\{1,2\}$, then the codebook size for all pairs of quantization indices is $L_1\times L_2$, i.e., a common codeword for each pair. Similar complexity holds for JD at the MCBS.

In this section, we propose a simpler scheme which involves 1) separate codeword transmission (i.e., $L_1+ L_2$ codewords) and 2) separate and sequential decoding. At the SCBS, the proposed scheme deploys TD transmission
 along with WZ binning for the quantization indices. While TD allows separate transmission and separate decoding, WZ binning allows sequential decoding \cite{hsce612}.
%
\begin{figure*}[t]
\normalsize
\renewcommand{\arraystretch}{0.55}
\begin{center}
\resizebox{\textwidth}{!}{%
\begin{tabular}{|c|c|c|c|c|c|c|c|}
\hline
\multicolumn{2}{|c|}{Block number}  & \multicolumn{2}{|c|}{$j$} & \multicolumn{2}{|c|}{$j+1$} & \multicolumn{2}{|c|}{$j+2$}\\
\hline
\multicolumn{2}{|c|}{UE1}& \multicolumn{2}{|c|}{$x_1(w_{1,j})$} & \multicolumn{2}{|c|}{$x_1(w_{1,j+1})$} & \multicolumn{2}{|c|}{$x_1(w_{1,j+2})$}\\
\hline
\multicolumn{2}{|c|}{UE2} & \multicolumn{2}{|c|}{$x_2(w_{2,j})$} & \multicolumn{2}{|c|}{$x_2(w_{2,j+1})$} & \multicolumn{2}{|c|}{$x_2(w_{2,j+2})$}\\
\hline
\multirow{5}{*}{\begin{minipage}[t]{0.06\columnwidth}%
Small
\newline
cell
\end{minipage}}& \multirow{2}{*}{Rx1} &\multicolumn{2}{|c|}{$\tilde{y}_{r1,j}\rightarrow \hat{y}_{r1,j}\rightarrow$} &  \multicolumn{2}{|c|}{$\tilde{y}_{r1,j+1}\rightarrow \hat{y}_{r1,j+1}\rightarrow$} &  \multicolumn{2}{|c|}{$\tilde{y}_{r1,j+2}\rightarrow \hat{y}_{r1,j+2}\rightarrow$} \\ \hhline{~~~~~}
& & \multicolumn{2}{|c|}{$l_{1,j}\rightarrow b_{1,j}$} & \multicolumn{2}{|c|}{$ l_{1,j+1}\rightarrow b_{1,j+1}$} &  \multicolumn{2}{|c|}{$l_{1,j+2}\rightarrow b_{1,j+2}$} \\
\hhline{~-------}
&\multirow{2}{*}{Rx2} &\multicolumn{2}{|c|}{$y_{r2,j}\rightarrow \hat{y}_{r2,j}\rightarrow$} &  \multicolumn{2}{|c|}{$\tilde{y}_{r2,j+1}\rightarrow \hat{y}_{r2,j+1}\rightarrow$} &  \multicolumn{2}{|c|}{$\tilde{y}_{r2,j+2}\rightarrow \hat{y}_{r2,j+2}\rightarrow$} \\ \hhline{~~~~~}
& & \multicolumn{2}{|c|}{$l_{2,j}\rightarrow b_{2,j}$} & \multicolumn{2}{|c|}{$ l_{2,j+1}\rightarrow b_{2,j+1}$} &  \multicolumn{2}{|c|}{$l_{2,j+2}\rightarrow b_{2,j+2}$} \\
\hhline{~-------}
&Tx &$x_{r1,j}(b_{1,j-1})$ & $x_{r2,j}(b_{2,j-1})$ & $x_{r1,j+1}(b_{1,j})$ &$x_{r2,j+1}(b_{2,j})$ & $x_{r1,j+2}(b_{1,j+1})$ & $x_{r2,j+2}(b_{2,j+1})$ \\
\hline
\multirow{4}{*}{\begin{minipage}[t]{0.06\columnwidth}%
Macro
\newline
cell
\end{minipage}}& Rx1  & \multicolumn{2}{|c|}{$\tilde{y}_{d1,j}$} &  \multicolumn{2}{|c|}{$\tilde{y}_{d1,j+1}$} & \multicolumn{2}{|c|}{$\tilde{y}_{d1,j+2}$} \\
\hhline{~-------}
 &Rx2  &  \multicolumn{2}{|c|}{$\tilde{y}_{d2,j}$} &  \multicolumn{2}{|c|}{$\tilde{y}_{d2,j+1}$} & \multicolumn{2}{|c|}{$\tilde{y}_{d2,j+2}$}\\
\hhline{~-------}
 &Rxr  &  $ \tilde{\mathbf{y}}_{dr1,j}$ &  $ \tilde{\mathbf{y}}_{dr2,j}$ &  $\tilde{\mathbf{y}}_{dr1,j+1}$ & $\tilde{\mathbf{y}}_{dr2,j+1}$ & $\tilde{\mathbf{y}}_{dr1,j+2}$ & $\tilde{\mathbf{y}}_{dr2,j+1}$ \\
\hhline{~-------}
 &\multirow{4}{*}{Dec.}  &$ \tilde{\mathbf{y}}_{dr1,j}\rightarrow $ & $\tilde{\mathbf{y}}_{dr2,j}\rightarrow$ &  $ \tilde{\mathbf{y}}_{dr1,j+1}\rightarrow$ & $\tilde{\mathbf{y}}_{dr2,j+1}\rightarrow$  &  $ \tilde{\mathbf{y}}_{dr1,j+2}\rightarrow$ & $\tilde{\mathbf{y}}_{dr2,j+2}\rightarrow $\\
 \hhline{~~~~~~~~}
 & & $ \tilde{b}_{1,j-1}$ & $ \tilde{b}_{2,j-1}$ &  $\tilde{b}_{1,j}$ & $\tilde{b}_{2,j}$  &  $ \tilde{b}_{1,j+1}$ & $\tilde{b}_{2,j+1}$\\
 \hhline{~~------}
& & \multicolumn{2}{|c|}{$ \tilde{y}_{d1,j-1}\rightarrow \tilde{l}_{1,j-1}\rightarrow \tilde{w}_{1,j-1}$} & \multicolumn{2}{|c|}{$  \tilde{y}_{d1,j}\rightarrow \tilde{l}_{1,j}\rightarrow \tilde{w}_{1,j}$} &  \multicolumn{2}{|c|}{$\tilde{y}_{d1,j+1}\rightarrow \tilde{l}_{1,j+1}\rightarrow \tilde{w}_{1,j+1}$} \\
\hhline{~~~~~}
& & \multicolumn{2}{|c|}{$ \tilde{y}_{d2,j-1}\rightarrow \tilde{l}_{2,j-1}\rightarrow \tilde{w}_{2,j-1}$} & \multicolumn{2}{|c|}{$  \tilde{y}_{d2,j}\rightarrow \tilde{l}_{2,j}\rightarrow \tilde{w}_{2,j}$} &  \multicolumn{2}{|c|}{$\tilde{y}_{d2,j+1}\rightarrow \tilde{l}_{2,j+1}\rightarrow \tilde{w}_{2,j+1}$}
\\
\hline
\end{tabular}}
\\
\vspace*{2mm}
{\small Table I: The encoding and decoding of the QF-WZTD scheme for massive MIMO HetNet.}\\
\end{center}
\vspace*{-8mm}
\end{figure*}
\vspace*{-4mm}
\subsection{The QF-WZTD Transmission Scheme}\label{sec: sss4}
 Table I helps describe the QF-WZTD transmission scheme where each UE transmission is identical to that in Section \ref{sec: simpsch}, but the SCBS transmission and MCBS decoding are different.
\subsubsection{At the SCBS: WZ binning and TD transmission}\label{subsec1}
\begin{itemize}[leftmargin=*]
  \item First, the SCBS deploys WZ binning, where it partitions the quantization indices for each UE data stream into equal-size bins \cite{hsce612}. That is, after obtaining the quantization indices $(l_{1,j},l_{2,j})$ as in Section \ref{sec: simpsch}, the SCBS finds the
two binning indices $b_{1,j}$ and $b_{2,j}$ that include $l_{1,j}$ and $l_{2,j}$, respectively.
  \item Second, the SCBS deploys TD transmission, where it generates separate codewords $U_{r1}$ and $U_{r2}$ for binning indices $b_{1,j}$ and $b_{2,j}$, respectively,  and transmits them in block $j+1$ in two separate phases in the block of duration $\beta_1$ and $\beta_2$, respectively. Therefore, in transmission block $j$, the SCBS generates its signals for forwarding as follows:
      \begin{align}\label{SCBT}
      \!\!\!\!\!\!\!\!\text{Phase $k$:}\;\mathbf{x}_{rk,j}=&\;\sqrt{\rho_{rk}/(\beta_kN)}\mathbf{U}_{rk}(b_{k,j-1}),\;k\in\{1,2\}
      \end{align}
      where $\beta_1+\beta_2=1$ and $\rho_{r1}+\rho_{r2}=P_r$. Note that the SCBS also deploys power control for transmitting $\mathbf{x}_{rk,j}$ with power $(\rho_{rk}/ \beta_k)$ in phase $k$.
\end{itemize}
\subsubsection{At the MCBS: separate and sequential decoding}\label{subsec2}
The MCBS performs the sliding window decoding to separately and sequentially decode each
bin index, quantization index and finally  the message  of each UE. Specifically, after ZF detection in (\ref{gaumodzf}), the  signals received over two phases in block $j+1$ from the SCBS are given by:
\begin{align}\label{UYDR}
\text{Phase $k$:}\;\tilde{\mathbf{y}}_{drk,j+1}&=\mathbf{x}_{rk,j+1}+\mathbf{A}_{dr,j+1}^{H}\mathbf{z}_{dk,j+1},\;k\in\{1,2\} 
\end{align}
Decoding for both UEs is done in a similar approach. For $\text{UE}_1$, the MCBS uses $\tilde{\mathbf{y}}_{dr1,j+1}$ in (\ref{UYDR}) and $\tilde{y}_{d1,j}$ in (\ref{gaumodzf}) to sequentially decode $1)$ the bin index $\hat{b}_{1,j}$ using $\tilde{\mathbf{y}}_{dr1,j+1}$, $2)$ the quantization index $\hat{l}_{1,j}$ using $\tilde{y}_{d1,j}$ given that $\hat{l}_{1,j}\in \hat{b}_{1,j}$, and then $3)$ $\text{UE}_1$ message $\hat{w}_{1,j}$ using $\tilde{y}_{d1,j}$ and $\hat{y}_{r1,j}(\hat{l}_{1,j})$.
\begin{rem}\label{DECSSD}
 With separate and sequential decoding for the bin index, quantization and then the UE message, each decoding step is similar to the decoding procedure in the point-to-point communication. This simpler procedure makes  the QF-WZTD scheme more attractive for a practical deployment.
\end{rem}
\vspace*{-4mm}
\subsection{Achievable Rate Region}
The achievable rate region by the QF-WZTD scheme is described below.
\begin{thm}\label{cr1}
For the considered massive MIMO HetNet, 
the rate region achieved by the QF-WZTD scheme
consists of all rate pairs $(R_1, R_2)$ satisfying:
\begin{align}\label{rc1}
\!\!\!\!&R_k\leq {\cal C}\left(\frac{P_k(M-N)}{d_{dk}^{\alpha}}+\frac{P_k}{(d_{rk}^{\alpha}/N)+Q_k}\right),\;
\text{s.t.}\;R_{bk}\leq \beta_k N{\cal C}\left(\frac{\rho_{rk}(M-N)}{\beta_kNd_{dr}^{\alpha}}\right),
\end{align}
for $k\in\{1,2\}$ where 
\begin{align}\label{rtb}
& R_{bk}={\cal C}\left(\frac{1}{Q_k}\Big[\frac{d_{rk}^{\alpha}}{N}+\frac{P_k}{1+(P_k(M-N)/d_{dk}^{\alpha})}\Big]\right),
\end{align}
for all $\beta_1,$ $\beta_2,$ $\rho_{r1}$ and $\rho_{r2}$ satisfying $\beta_1+\beta_2=1$ and $\rho_{r1}+\rho_{r2}=P_r$. The transmission rates for the quantization indices ($R_{q1}$ and $R_{q2}$) are given as in (\ref{qrt1}). 
\end{thm}
\IEEEproof
Results in (\ref{rc1}) are
 obtained considering   the signaling in Section \ref{subsec1} and the error analysis for decoding rule in   \ref{subsec2}. The proof is similar to the QF-WZ scheme for the relay channel with three sequential decoding steps in \cite{hsce612} but considering the phase duration lengths for the signals sent by the SCBS. We omit the detailed proof because of space limitation.
\endIEEEproof
Comparing  with Theorem \ref{thss}, we have the following comments on Theorem \ref{cr1}:
\begin{itemize}[leftmargin=*]
  \item Since the QF-WZTD scheme deploys separate decoding, the rate region in (\ref{rc1}) consists of individual rate constraints only, without the sum rate constraint as in (\ref{dmcac}) under  the QF-JD scheme, which deploys joint decoding.
  \item Despite the absence of the sum rate constraint, the two UEs cannot simultaneously transmit at their maximum rate since they share the SCBS-MCBS link through TD  transmission.
  \item From (\ref{qrt1}) and (\ref{rtb}), the transmission rate of each binning index is less than that of the quantization index $(R_{bk}\leq R_{qk})$, which is expected. By definition,  each bin index represents a group of quantization indices.
  \item In the QF-WZTD  scheme, while each decoding step is similar to that in the point-to-point communication (Remark \ref{DECSSD}), in addition to the quantization noise variances $Q_1$ and $Q_2$, more parameters need to be optimized  than the QF-JD scheme. These parameters include the phase durations $(\beta_1,\beta_2)$ and power allocation $(\rho_{r1},\rho_{r2})$.
  \item  Because of the orthogonality property of massive MIMO \cite{lars} and TD transmission with separate decoding, the transmission and decoding for each UE message is similar to the basic single UE relay channel in \cite{hsce612}. However, as all UEs share the same relaying link from the SCBS to the MCBS,  quantization at the SCBS is not simply optimized by considering multiple orthogonal single user relay channels. Further details are given in Section \ref{comps2}.
  \end{itemize}
  Next, we show that the QF-WZTD scheme in fact achieves the same rate region as the QF-JD scheme with a lower complexity, despite having more optimization parameters.
\vspace*{-3mm}
\section{Comparison of QF-WZTD and QF-JD schemes.}\label{comps2}
We now compare the two proposed schemes in terms of their achievable rate regions, computational complexities and decoding delays. For both schemes, the
decoding delays are identical as the MCBS waits one block before it starts decoding the messages of different UEs. Hence, in the following, we focus on    the
achievable rate region  and computational complexity.
\vspace*{-4mm}
\subsection{Rate Region}
In general, simple sequential and separate decoding used in the QF-WZTD scheme results in  a smaller rate region than that under the JD used in the QF-JD scheme \cite{hsce612}. However, for the considered HetNet, they have the same rate region. 
\vspace*{-2mm}
\begin{thm}\label{equr}
For the considered massive MIMO HetNet, the achievable rate region by the QF-WZTD scheme is identical to that of the QF-JD scheme.
\end{thm}
\vspace*{-3mm}
\IEEEproof
  Both schemes achieve the same weighted sum rate since the two constraints in (\ref{rc1}) lead to the same $Q_1^{\ast}$ and $Q_2^{\ast}$ in (\ref{opsI1}) except $\lambda_k$ is replaced by $\lambda_{sk}$ for $k\in\{1,2\}$ where:
  \begin{align}\label{la222}
    \lambda_{sk}= \left(1+\frac{\rho_{rk}(M-N)}{\beta_kNd_{dr}^{\alpha}}\right)^{\beta_k N}.
  \end{align}
  Hence, the two schemes achieve the same maximum weighted sum rate when $\lambda_{sk}=\lambda_k^{\ast}$ in (\ref{opsI1}), which is possible by setting:
  \begin{align}\label{EXOP}
\rho_{rk}^{\ast}=\beta_k^{\ast}P_r,\;\beta_k^{\ast}=\log(\lambda_k^{\ast})/\log(\lambda_s),\;k\in\{1,2\},
\end{align}
where (\ref{EXOP}) satisfies the conditions in (\ref{rc1}).
\endIEEEproof
\begin{rem}\label{remm2}
  Theorems \ref{cr1} and \ref{equr} have the following implications.
  \begin{itemize}[leftmargin=*]
    \item Although the QF-WZTD scheme requires WZ binning and TD transmission from the SCBS, it in fact requires no additional parameter optimization than those in the QF-JD scheme. This is because the optimal phase durations $(\beta_1,\beta_2)$ and power allocation $(\rho_{r1},\rho_{r2})$ are conveniently  obtained as functions of the optimal quantization $(Q_1,Q_2)$ for the QF-JD scheme as shown in (\ref{la222}) and (\ref{EXOP}). Moreover, by substituting $\rho_{rk}^{\ast}$ in (\ref{EXOP}) into (\ref{rc1}), it is optimal that the SCBS transmits each bin index with the same power $P_r$ in each phase. Hence, there is no need for different power allocation in each transmission phase at the SCBS.
    \item As the QF-WZTD scheme has the same rate performance as the QF-JD, the same rate is achieved by the sole deployment of WZ binning (QF-WZ) or TD (QF-TD), but with higher complexity than the QF-WZTD scheme, as discussed next. 
  \end{itemize}
\end{rem}
\vspace*{-5mm}
\subsection{Complexity}\label{sec: COMP}
 Although the SCBS's transmission in the QF-WZTD scheme requires optimization of the phase durations and power allocations, Remark \ref{remm2} shows that it can be done easily and hence incur negligible computational complexity. Moreover, although WZ binning is an extra encoding step at the SCBS, only minor computation is needed for sorting the quantization indices into equal size groups. Therefore, the complexity comparison between the two schemes is determined by the codebook size  and the decoding complexity of each scheme.
\subsubsection{Codebook size}
Both schemes have the same codeword generation at each UE and the same estimation of the quantization indices at the SCBS. Hence, we only focus on the codebook size for the signals transmitted by the SCBS. Let $n$  be the length of the transmitted codewords. In the QF-JD scheme,  $2^{n(R_{q1}+R_{q2})}$
  common codewords are generated to represent all pairs of quantization indices. However, in the QF-WZTD scheme,  $2^{nR_{b1}}+2^{nR_{b2}}$ separate codewords are generated to represent all binning indices.
Since $R_{bk}<R_{qk},$ for $k\in\{1,2\}$,  the number of binning indices is smaller than that of the quantization indices. It is clear that the QF-WZTD scheme requires much fewer codewords than the QF-JD scheme. For example, with $R_{q1}=R_{q2}=1$ bps/Hz and $R_{b1}=R_{b2}=0.5$ bps/Hz, only $2\cdot2^{0.5n}$ coderwords are generated for the QF-WZTD scheme, instead of $2^{2n}$ codewords for the QF-JD scheme.
%
%
%
\subsubsection{Decoding complexity}
  At the MCBS, considering an ML decoding, we define
the decoding complexity as the number of likelihoods calculated to estimate
the transmitted messages, which are given as follows.
In the QF-JD scheme, the MCBS jointly decodes both UE messages and quantization indices by calculating $2^{n(R_1+R_{q1}+R_2+R_{q2})}$ likelihoods. However, in the QF-WZTD scheme, the MCBS separately and sequentially decodes each  bin index,  quantization index within the decoded bin index, and then  UE message, by calculating $2^{nR_{bk}},$ $2^{n(R_{qk}-R_{bk})},$ and $2^{nR_k}$ likelihoods, respectively, for $k\in\{1,2\}$. Hence, the total likelihoods for calculation are: $2^{nR_1}+2^{n(R_{q1}-R_{b1})}+2^{nR_{b1}}+$ $2^{nR_2}+2^{n(R_{q2}-R_{b2})}+2^{nR_{b2}}$.
Similar to the  codebook size, the decoding of the QF-WZTD scheme is much simpler than the QF-JD scheme. For example, with  $R_{1}=R_{2}=1.5$ bps/Hz, $R_{q1}=R_{q1}=1$ bps/Hz and $R_{b1}=R_{b2}=0.5$ bps/Hz, the decoding complexity for the QF-WZTD scheme is $2(2^{1.5n}+2\cdot2^{0.5n}),$ while it is $2^{5n}$ for the QF-JD scheme.
\begin{rem}\label{remRC}
  Based on Theorem \ref{equr}, Remark \ref{remm2}, and Section \ref{sec: COMP}, the QF-WZTD scheme is the preferred scheme: It  achieves the rate region of the QF-JD scheme with a much lower complexity.
\end{rem}
\subsubsection{Impact of deploying TD with WZ binning}
Remark \ref{remm2} clarifies that the sole deployment of WZ binning  or TD  leads to the same rate region of the QF-WZTD scheme but with higher complexity. This  difference is explained below. \\
In the QF-WZ scheme (without TD transmission):
\begin{itemize}[leftmargin=*]
  \item The SCBS generates a common codeword for each pair of binning indices for $\text{UE}_1$ and $\text{UE}_2$, resulting in a total of $2^{n(R_{b1}+R_{b2})}$ codewords, which is higher than $2^{nR_{b1}}+2^{nR_{b2}}$ separate codewords for the QF-WZTD scheme.
  \item The MCSB sequentially decodes 1) both binning indices (jointly) by calculating $2^{n(R_{b1}+R_{b2})}$ likelihoods, 2) each quantization index, and 3) each UE message  as in the QF-WZTD scheme. Clearly, the first step requires more calculations than the QF-WZTD scheme since $2^{n(R_{b1}+R_{b2})}>2^{nR_{b1}}+2^{nR_{b2}}$, which shows the benefit of separate decoding facilitated by TD transmission.
\end{itemize}
In the QF-TD scheme (without WZ binning):
\begin{itemize}[leftmargin=*]
  \item The SCBS will generate a separate codeword for each quantization index, resulting in a total of   $2^{nR_{q1}}+2^{nR_{q2}}$ codewords, which is higher than $2^{nR_{b1}}+2^{nR_{b2}}$ codewords of the QF-WZTD scheme as $R_{q1}>R_{b1}$ and $R_{q2}>R_{b2}$.
  \item The MCSB can sequentially and separately decode each quantization index by calculating $2^{nR_{qk}}$ likelihoods for $k\in\{1,2\},$ and then  each UE message  as in the  QF-WZTD scheme.  However, the first step of the QF-TD scheme requires more calculations than the first two steps of the QF-WZTD scheme since $2^{nR_{qk}}>2^{n(R_{qk}-R_{bk})}+2^{nR_{bk}}$ as $R_{qk}>R_{bk}$. 
\end{itemize}
Therefore, both WZ binning and TD transmission are needed to simplify the transmission design.
\vspace*{-5mm}
\section{Generalization to $K$ UEs}\label{sec: gen}
In previous sections, we focused on
 the two-UE case for the QF relaying design in a massive MIMO HetNet. Here, we generalize our results to the $K$-UE case.
 Let $\mathcal{K}=\{1,2,\ldots, K\}$, then $P_k,$ $d_{dk},$ $d_{rk},$ and $Q_k$ are $\text{UE}_k$'s transmit power, its distances to the MCBS and the SCBS, and the quantization noise variance of
its data stream at the SCBS, respectively.
\vspace*{-5mm}
\subsection{The QF-JD Scheme}
The transmission scheme for $K$ UEs is  similar to that of two UEs in Section \ref{secsigsc}.
Each UE transmits a new message in each block, the SCBS deploys ZF detection,
quantizes the  signal received from each UE, and sends a common codeword for all quantization
indices in the next block. The MCBS decodes all messages jointly for some quantization indices over two
consecutive blocks (sliding window decoding).
\subsubsection{Achievable rate region}
Using the similar encoding and decoding techniques for  the two-UE case in Section \ref{secsigsc}, we obtain
the following rate region for $K$-UE transmission:
\begin{thm}\label{cr4}
For $K$-UE massive MIMO HetNet, the achievable rate region of the QF-JD scheme consists of all $K$-tuples rate vectors $(R_1, R_2,\ldots, R_K)$ satisfying:
\begin{align}\label{rc4}
R_k&\leq \min\{I_k,J_k\},\; k\in\mathcal{K},\quad
\sum_{k\in\Lambda}R_k\leq J_{\Lambda},\;\forall\;\Lambda \subseteq \mathcal{K},
\end{align}
where $\Lambda$ is a subset of $\mathcal{K}$ and
\begin{align}\label{rcs4}
I_k=&\;{\cal C}\left(\frac{P_k(M-N)}{d_{dk}^{\alpha}}+\frac{P_k}{(d_{rk}^{\alpha}/N)+Q_k}\right),\;J_k={\cal C}\left(\frac{P_k(M-N)}{d_{dk}^{\alpha}}\right)+\xi,\\
J_{\Lambda}=&\;\xi+\sum_{k\in \Lambda}{\cal C}\left(\frac{P_k(M-N)}{d_{dk}^{\alpha}}\right),\;
\xi= \;N{\cal C}\left(\frac{P_r(M-N)}{Nd_{dr}^{\alpha}}\right)
-\sum_{k=1}^{K}{\cal C}\left(\frac{d_{rk}^{\alpha}}{NQ_k}\right).\nonumber
\end{align}
  \noindent The transmission rate $R_{qk}$ for the quantization index $l_k$ is given by (\ref{qrt1}) where $k\in \mathcal{K}$.
\end{thm}
\IEEEproof
The proof is similar to the proof of Theorem \ref{thss}  and is omitted to avoid repetition.$\blacksquare$\\
 Theorem \ref{thss} is a special case of the general result in Theorem \ref{cr4}. With $K=2$ and all possible subsets $\mathcal{S}=\{1\},\{2\},\{1,2\}$, we arrive the rate region for the two-UE case.
%
\subsubsection{Optimal quantization}
As in Section \ref{OPQF}, the optimal quantization at SCBS is specified by the quantization noise variances $(Q_1,Q_2,\ldots,Q_K)$ that maximize the weighted sum rate. Hence, the optimization problem can be formulated as:
%
\begin{align}\label{WSROK}
\max_{\substack{\{Q_k\}}} &\sum_{k=1}^{K}\mu_k R_k\quad
\text{s.t.}\;  R_k\leq \min\{I_k,J_k\},\; Q_k\geq 0,\;  k \in \mathcal{K},\;
\sum_{k\in\Lambda}R_k\leq J_{\Lambda},\;\forall\;\Lambda \subseteq \mathcal{K}, 
\end{align}
where $I_k,$ $J_k,$  and $J_{\Lambda}$ are given in Theorem  \ref{cr4}, and  $\mu_k\in[0,1]$ is the weight for $\text{UE}_k$ with $\sum_{k=1}^{K}\mu_k=1$. Define:
%
\begin{align}\label{order}
\omega_k&=\frac{a_k+b_k}{\mu_k b_k},\;\text{where}\;a_k=1+P_k(M-N)/d_{dk}^{\alpha},\;b_k=P_kN/d_{rk}^{\alpha}.
\end{align}
 Without loss of generality, assume that $\omega_1\leq \omega_2\leq \ldots\leq \omega_K$.
Then, the solution for problem (\ref{WSROK}) is given as follows:
\vspace*{-2mm}
\begin{thm}\label{cor1K}
The optimal $Q_k^{\ast},$ $k\in\{1,2,\ldots,K\},$ for problem (\ref{WSROK}) are given as in (\ref{opsI1}),
but with $\lambda_k^{\ast}$ obtained as follows:
\begin{align}\label{FFL1}
\lambda_k^{\ast}=\left\{\begin{array}{cl}
  \frac{\mu_k b_k}{a_k}\left(x_s+\omega_{\upsilon^{\ast}}-\frac{1}{\mu_k}\right)& \text{for}\;k\in\{1,\ldots,\upsilon^{\ast}\} \\
 1  & \text{for}\;k\in\{\upsilon^{\ast}+1,\ldots,K\}
\end{array}\right.,
\end{align}
where $x_s$ is the unique positive real root of the polynomial function $f(x_s)$ given by:
\begin{align}\label{fxsp}
f(x_s)=\prod_{k=1}^{\upsilon^{\ast}}\left(x_s+\omega_{\upsilon^{\ast}}-\frac{1}{\mu_k}\right)-
\lambda_s\prod_{k=1}^{\upsilon^{\ast}}\frac{a_k}{\mu_kb_k},
\end{align}
 and $\upsilon^{\ast}$ is the maximum $\upsilon \in\{1,2,\ldots,K\}$ such that:
\begin{align}\label{finth}
\prod_{k=1}^{\upsilon^{\ast}}\left(\omega_{\upsilon^{\ast}}-\frac{1}{\mu_k}\right)\leq \lambda_s\prod_{k=1}^{\upsilon^{\ast}}\frac{a_k}{\mu_kb_k}.
\end{align}
\end{thm}
\IEEEproof
Following similar steps in the proof of Theorem \ref{cor1} until (\ref{EWSMUU}) in step $4$ which can be reexpressed for the $K$-UE case as follows:
 \begin{align}\label{EWSMUUK}
              \!\!\!&\min_{\substack{\{\lambda_k\}}}   \sum_{k=1}^{K}\mu_k\log\left(a_k+\frac{b_k}{\lambda_k}\right), \;\text{s.t.}\;\prod_{k=1}^{K}\lambda_k=\lambda_s,\;\lambda_k^{-1}\leq 1,\;\forall k \in \mathcal{K},
 \end{align}
      where $a_k$ and $b_k$ are given in (\ref{order}).
      The above optimization problem is non-convex, as the equality constraint is not affine. However, we now show that it can be transformed to a convex problem. 
   Following similar transformation of a GP to a convex problem \cite{BYDC}, define
  the  new variables $\delta_k=\ln \lambda_k$ for $k\in\{1,2,\ldots,K\}$ and take the logarithms of the constraints. Then, the problem in (\ref{EWSMUUK}) becomes:
        \begin{align}\label{newop}
        \!\!\!&\min_{\substack{\delta_k, k\in\{1,2,\ldots,K\}}} \sum_{k=1}^{K}\mu_k \ln\big(a_k+b_ke^{-\delta_k}\big),
        \;\text{s.t.}\;-\ln(\lambda_s)+\sum_{k=1}^{K}\delta_k=0,\;\;-\delta_k\leq 0, \; k\in\mathcal{K}.
        \end{align}
  By applying the KKT conditions, we obtain $\lambda_k^{\ast}$ in (\ref{FFL1}) in a way similar (although not exact) to the water-filling approach for power allocation in the parallel MIMO channel \cite{hsce612}, where $x_s+\omega_{\upsilon^{\ast}}$ represents the water-level of the optimal quantization.  We determine $x_s+\omega_{\upsilon^{\ast}},$ with the help of the Descartes' rule of signs \cite{DCRS} as explained in Appendix B.  Note that we cannot use the typical bi-section search approach as used in the power allocation problem to find the water level, because the equality constraint in (\ref{newop})  is in the form of a sum of logarithms of optimization variables, instead of linear addition in the conventional water-filling problem.
\endIEEEproof
\begin{rem}
The optimal $\upsilon^{\ast}$ in (\ref{finth}) is obtained through multiple iterations, e.g. start with $\upsilon=K/2$ and then increase or decrease $\upsilon$ till reaching $\upsilon^{\ast}$ that satisfies (\ref{finth}). With initial  $\upsilon=K/2$, the maximum number of iterations is $K/2$.
\end{rem}
\begin{rem}
All comments in Remark \ref{fineQ} hold for Theorem \ref{cor1K}. Furthermore, Theorem \ref{cor1K} implies that the SCBS deploys a finer quantization for the UE$_k$ data stream as $\mu_k$ increases and  its channel to the SCBS becomes stronger compared to the MCBS. Furthermore, for a UE with a very small weight or a much weaker channel to the SCBS than to the MCBS, the SCBS does not quantize or forward the UE data,  and the MCBS decodes its message by only using the signal received from the UE over the direct link.
This implication resembles the power allocation in water-filling for the MIMO channel where more power is allocated to the stronger channels.
\end{rem}
\vspace*{-6mm}
\subsection{The QF-WZTD Scheme}
The QF-WZTD scheme for $K$ UEs is quite similar to that of two UEs in Section \ref{sec: sss4}.
Each UE transmits a new message in each block, the SCBS deploys ZF detection, quantizes each UE data stream and uses WZ binning for each UE quantized data stream.
Then, the SCBS  transmits $K$ separate codewords for the $K$ binning indices in the next block via TD transmission in $K$ phases (time slots) of durations $\beta_1,$ $\beta_2,$ $\ldots$ and $\beta_K$ where $\sum_{k=1}^{K}\beta_k=1$.

The MCBS performs sliding window decoding and decodes each UE message separately and sequentially as follows. To decode the $\text{UE}_k$ message, which is sent in block $j$,
the MCBS utilizes the signal received from the SCBS in block $j+1$ and phase $k$ to decode the binning index. Then, the MCBS utilizes the signal received directly from $\text{UE}_k$ in block $j$ to decode the quantization index and then the $\text{UE}_k$ message.
\subsubsection{Achievable rate region}
Following the similar analyses of the two-UE case in Section \ref{sec: sss4}, we obtain
the following rate region under the QF-WZTD scheme for $K$ UEs.
\begin{cor}\label{cr5}
For $K$-UE massive MIMO HetNet, the achievable rate region of the QF-WZTD scheme consists of all $K$-tuple rate vectors $(R_1, R_2,\ldots, R_K)$ satisfying the constraints in (\ref{rc1}) but for $k\in\{1,2,\ldots,K\}$
 where the power parameters satisfy the constraint $\sum_{k=1}^{K}\rho_{rk}=P_r$.
\end{cor}
\vspace*{-2mm}
\IEEEproof
The proof is similar to the proof of Theorem \ref{cr1} and is omitted.
\endIEEEproof
\subsubsection{Comparison with the QF-JD scheme}
By comparing the two schemes and using the same approach as in the proof of Theorem \ref{equr} and the complexity derivation in Section \ref{sec: COMP}, we obtain similar results to those in Section \ref{comps2} as in the following two corollaries:
\begin{cor}\label{cr6}
In a K-UE massive MIMO HetNet, the QF-WZTD scheme achieves the rate region as the QF-JD scheme with the same optimal parameters in Theorem \ref{equr} except the set $k$ is changed from $\{1,2\}$ to $\{1,2,\ldots,K\}$.
\end{cor}
\begin{cor}\label{cr7}
In a K-UE massive MIMO HetNet, the codebook size and the decoding complexities are defined and derived as in Section \ref{sec: COMP}.  For the QF-WZTD scheme, the codebook size is $\sum_{k=1}^{K}2^{nR_{bk}}$ while the decoding complexity is $\sum_{k=1}^{K}\big(2^{nR_k}+2^{n(R_{qk}-R_{bk})}+2^{nR_{bk}}\big)$. For the QF-JD scheme,  the codebook size is $2^{n\sum_{k=1}^{K}R_{qk}}$ while the decoding complexity is $2^{n\sum_{k=1}^{K}(R_k+R_{qk})}$.
\end{cor}
\vspace*{-3mm}
It is clear from Corollary \ref{cr7} that in the QF-WZTD scheme, the complexity increases linearly with the number of UEs, while it increases exponentially in the QF-JD scheme. As an example, with $R_{k}=1.5$ bps/Hz, $R_{qk}=1$ bps/Hz and $R_{bk}=0.5$ bps/Hz, for $k\in\{1,2,\ldots,K\}$, Table II shows the codebook size of the two schemes for different numbers of UEs. With $K$ UEs, the decoding complexity  is $K(2^{1.5n}+2\cdot2^{0.5n})$ for the QF-WZTD scheme and $(2^{2.5n})^K$ for the QF-JD scheme.
\vspace*{-7mm}
\begin{figure*}[t]
\normalsize
\renewcommand{\arraystretch}{0.75}
\begin{center}
\begin{tabular}{|c|c|c|c|c|c|}
\hline
Number of UEs  & $3$ & $5$ & $10$ & $K$\\
\hline
QF-WZTD scheme& $3\cdot2^{0.5n}$ & $5\cdot2^{0.5n}$ & $10\cdot2^{0.5n}$ & $K\cdot2^{0.5n}$\\
\hline
QF-JD scheme & $2^{3n}$ & $2^{5n}$ & $2^{10n}$ & $(2^n)^K$\\
\hline
\end{tabular}
\\
\vspace*{2mm}
{\small Table II: Codebook size of the transmitted signals at the SCBS in the QF-WZTD and QF-JD schemes.}\\
\end{center}
\vspace*{-12mm}
\end{figure*}
\vspace*{-5mm}
\section{Numerical Results}\label{sec:cap. gau}
We now provide numerical results for the rate region and the optimal quantization of the QF-JD and QF-WZTD schemes.
In the simulation, all UEs have the same power $P_k=P$ for $k\in\mathcal{K}$ while the SCBS power is $P_r=5P$ unless otherwise mentioned.
 The SCBS and MCBS have $50$ and $500$ antennas, respectively. Unless specified, the inter-node distances in meters are set to: $d_{d1}=105, d_{d2}=110,$ and $d_{dr}=100,$ while $d_{r1}$ and $d_{r2}$ are given in each figure. These distances are valid for $5$G systems with a small cell radius at the  order of $100$ m and a macro cell radius of 1 Km \cite{5GRD}. We set the pathloss exponent $\alpha = 2.7$. We set the received SNR at the MCBS  from $\text{UE}_1$ as follows: $\text{SNR}=10\log_{10}\left(P_1(M-N)d_{d1}^{-\alpha}\right).$
In all figures, we set $\text{SNR}=1$dB.
\vspace*{-4mm}
\subsection{Achievable Rate Regions}
\begin{figure}[!t]
\begin{minipage}[b]{0.48\linewidth}
    \includegraphics[width=0.95\textwidth]{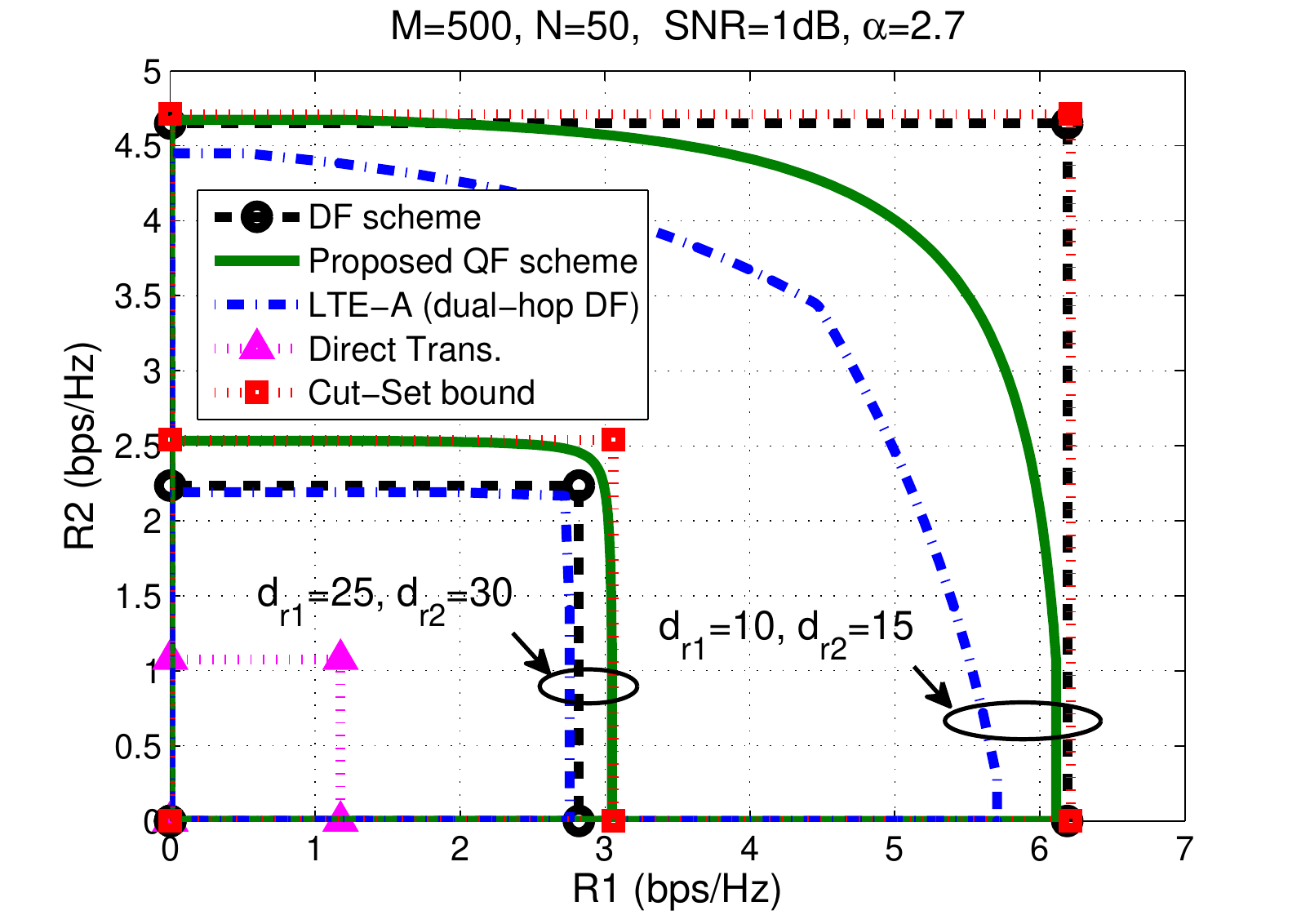}
    \caption{Rate regions  of the massive MIMO HetNet under with different transmission schemes.}
    \label{fig:asyro1}
    \end{minipage}
    \vspace*{-4mm}
    \hfill
\begin{minipage}[b]{0.48\linewidth}
    \includegraphics[width=0.95\textwidth]{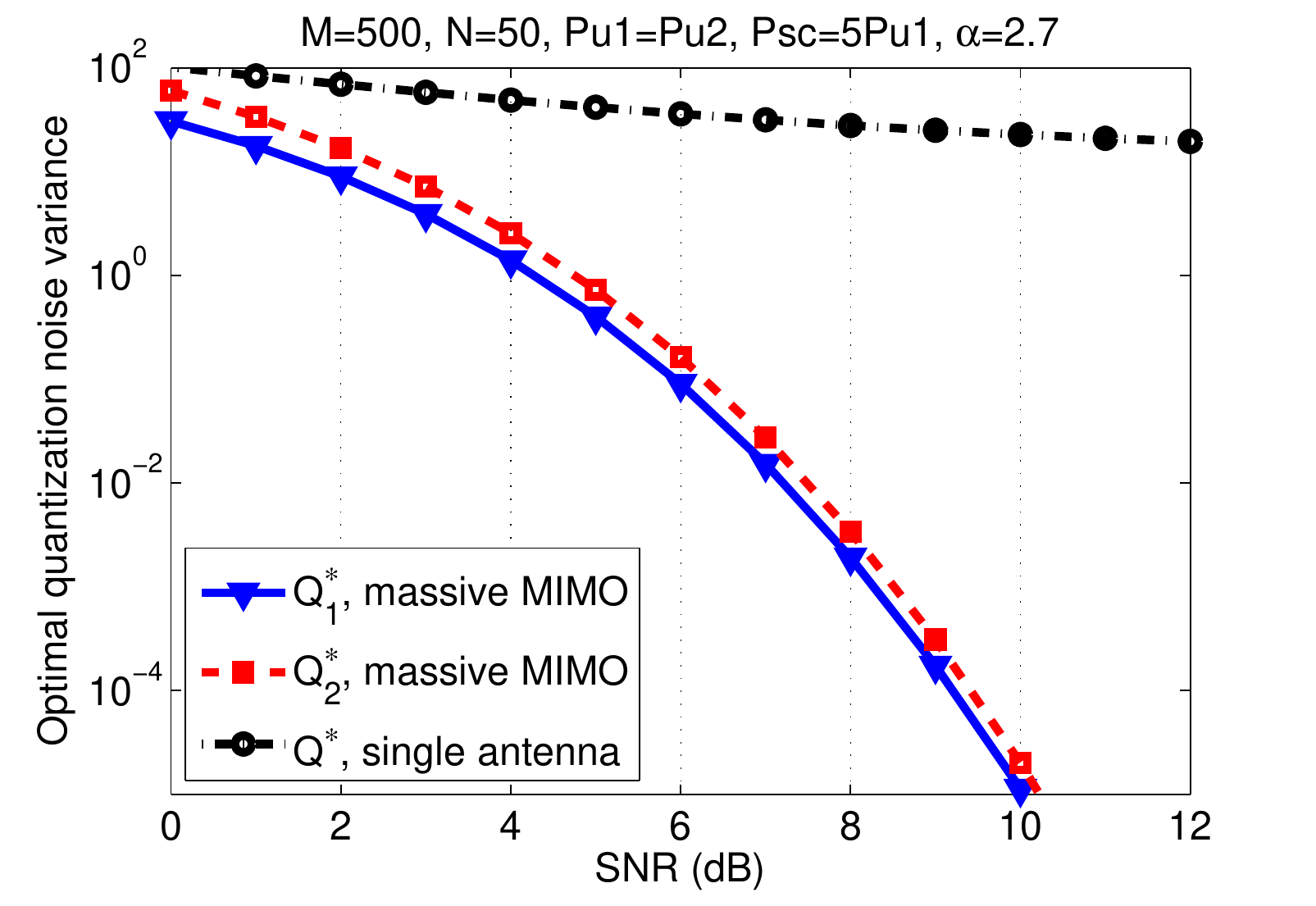}
    \caption{Optimal quantization at the small cell to maximize the sum rate.}
    \label{fig:asyro3}
    \end{minipage}
    \vspace*{-4mm}
\end{figure}

Fig. \ref{fig:asyro1} shows the rate region of the massive MIMO HetNet under the proposed QF schemes. (Note that by Theorem \ref{equr}, the identical rate region is achieved by the QF-WZTD scheme.) For comparison, we also plot the rate regions of several other schemes, including DF relaying \cite{MARC2}, LTE-A (dual-hop DF scheme), direct transmission (from UEs to the MCBS without the SCBS) and the cut-set outer bound \cite{hsce612}. For DF relaying, we apply the scheme in \cite{MARC2} to the channel model in (\ref{gaumodzf}). For the LTE-A scheme, we consider the two-hop transmission where both UEs transmit to the SCBS in phase $1$ and the SCBS forwards their information to the MCBS in phase $2$.  Fig. \ref{fig:asyro1} shows the rate regions under two sets of UE-to-SCBS distances $d_{r1}$ and $d_{r2}$   (near or far). Both  the QF and DF schemes outperform the direct transmission and LTE-A schemes, due to deploying the SCBS in full-duplex mode and utilizing direct links from the UEs to the MCBS. However, as expected from \cite{hsce612}, comparing DF and QF schemes, neither always outperforms the other: QF relaying outperforms DF relaying when the UEs are relatively further away from the SCBS, but underperforms  DF relaying in the opposite scenario.
%
\noindent
\noindent
\begin{figure}[!t]
\begin{minipage}[b]{0.48\linewidth}
    \includegraphics[width=0.95\textwidth]{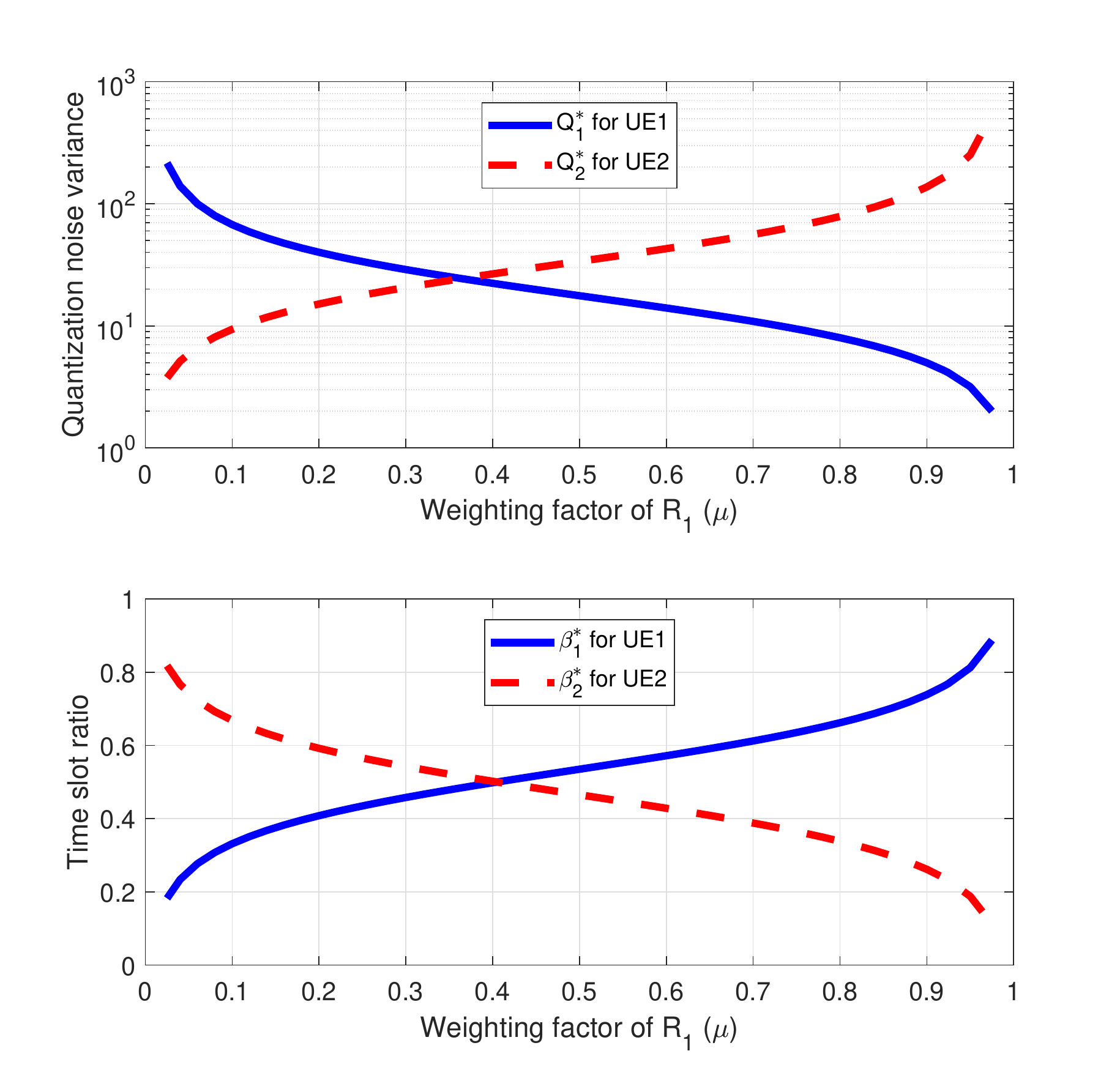}
    \caption{Optimal quantization noise variances and phase durations to maximize the weighted sum rate of the QF-WZTD scheme.}
    \label{fig:asyro5}
    \end{minipage}
    \vspace*{-5mm}
    \hfill
\begin{minipage}[b]{0.48\linewidth}
    \includegraphics[width=0.95\textwidth]{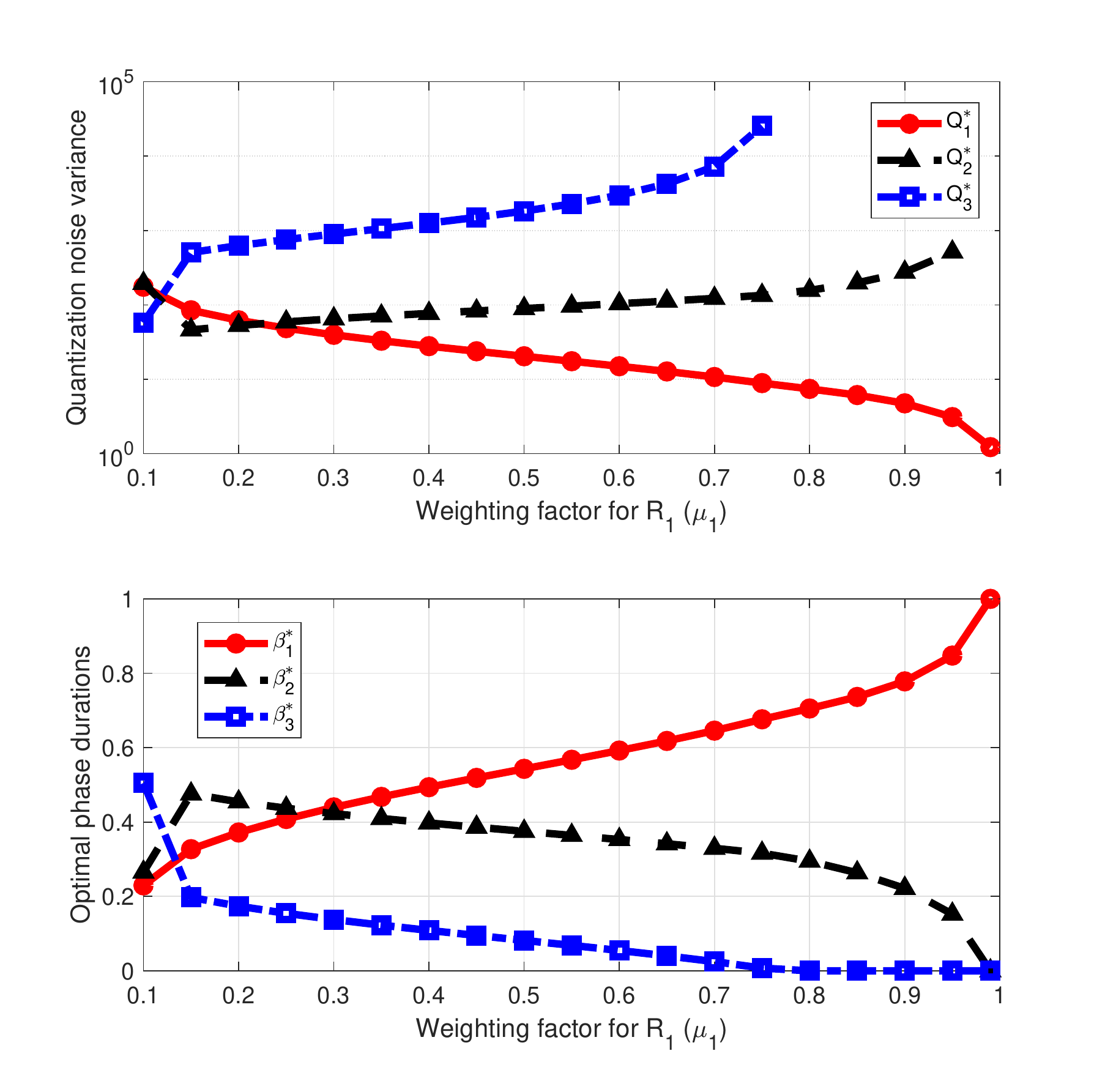}
    \caption{Optimal quantization noise variances and phase durations to maximize the weighted sum rate of \textbf{three UEs} with $\mu_2=0.75(1-\mu_1)$ and $\mu_3=0.25(1-\mu_1)$.}
    \label{fig:asyro7}
    \end{minipage}
    \vspace*{-5mm}
\end{figure}
\vspace*{-4mm}
\subsection{Optimal Quantization}
Fig. \ref{fig:asyro3} shows the optimal quantization noise variances $(Q_1^{\ast}, Q_2^{\ast})$ that maximize the sum rate  of massive MIMO and single antenna HetNets.  We set the SCBS power $P_{sc} = 5P$. As in Remark \ref{fineQ}, for the massive MIMO HetNet, $Q_1^{\ast}$ and $Q_2^{\ast}$ decrease much more significantly with SNR, as compared with those in the single antenna HetNet. Hence, with massive MIMO, the SCBS can send clear (quantized) versions of its received signals to the MCBS.

Fig. \ref{fig:asyro5} shows the optimal quantization noise variances $(Q_1^{\ast},Q_2^{\ast})$ and phase durations $(\beta_1^{\ast},\beta_2^{\ast})$ at different values of weighting factor $\mu_1$ for the weighted sum rate maximization under both QF-JD and QF-WZTD schemes. We set $d_{r1}=25$  and $d_{r2}=30$.
 As the weighting factor $\mu_i$ of each UE rate increases, the SCBS performs a finer quantization to send a clearer version of that UE signal to the MCBS. Similar results hold for the optimal phase durations where as  $\mu_1$,  the SCBS sends the quantization index of $\text{UE}_1$ over a longer phase duration to increase its rate.
\vspace*{-4mm}
\subsection{$K$-UE Transmission}
For the $K$-UE case, we obtain similar results to Fig. \ref{fig:asyro5}  but for $3$ UEs with the following distances: $d_{dr}=100,$ $d_{d1}=105,$ $d_{d2}=110,$ $d_{d3}=120,$ $d_{r1}=30,$ $d_{r2}=40,$ and $d_{r3}=50$. The results in Fig. \ref{fig:asyro7} are obtained versus $\mu_1$ where $\mu_2=0.75(1-\mu_1)$ and $\mu_3=0.25(1-\mu_1)$. The results are quite similar to Fig. \ref{fig:asyro5}. However, for $\mu_1\geq 0.8$, so that $\mu_3\leq0.05,$  since $\text{UE}_3$ has the weakest link to the SCBS, the optimal quantization $Q_3^{\ast}$ is very high (approaches $\infty$), i.e., the SCBS does not QF $\text{UE}_3$'s signal and the MCBS decodes it through the direct link only. Similarly for the optimal phase durations $\beta_1^{\ast},\beta_2^{\ast},$ and $\beta_3^{\ast}$, for $\mu_1\geq0.8$, $\beta_3^{\ast}=0$ since the SCBS  does not QF $\text{UE}_3$'s signal.
\vspace*{-4mm}
\section{Conclusion}\label{sec:conclusion}
We have investigated how massive MIMO impacts the uplink transmission design for  the HetNet with ZF detection at the SCBS and MCBS, while using QF relaying at the SCBS. For rate maximization,  we derived the optimal quantization parameters
whose size was shown to be equal to the number of UEs instead of antennas, as in traditional MIMO systems, and
obtained their values as explicit functions of the large-scale fading.
Results show that a
 finer quantization of UE data streams is performed when the UE-SCBS links become stronger as compared with the UE-MCBS links.
For the proposed QF-WZTD scheme with WZ binning and multiple-timeslot transmission at the SCBS, the MCBS can deploy separate and sequential decoding for each UE message. Consequently, the codebook size and decoding complexity grow linearly with the number of UEs instead of exponentially as in common transmission and joint decoding. Despite the simplicity, we show that there is no loss to the rate region while the time slot durations are conveniently optimized based on the  optimal quantization parameters.
\appendices
\vspace*{-4mm}
\section*{Appendix A: Proof of Theorem \ref{thss}}
The discrete memoryless MARC with orthogonal receivers (as in (\ref{gaumodzf})) is specified by a collection of probability mass functions (pmfs) $p(\tilde{\mathbf{y}}_{dr}|\mathbf{x}_r)$$p(\tilde{y}_{d1}|x_1)p(\tilde{y}_{d2}|x_2)p(\tilde{y}_{r1}|x_1)p(\tilde{y}_{r2}|x_2)$.  
The $(2^{nR_1} , 2^{nR_2}, n, Pe)$ code follows the standard definitions in \cite{hsce612}.
We consider $B$ independent transmission blocks each of length $n$. Two sequences of $B-1$ messages $w_{1,j}$ and $w_{2,j}$ for $j\in[1:B-1]$ are sent in $nB$ transmissions. Hence, both UEs do not transmit in the last block ($B$) which reduces the rates by a factor of $1/B$ (negligible as $B\rightarrow \infty$) \cite{hsce612}.
\subsubsection{Codebook generation}
The codebook generation can be explained as follows. First, fix the pmf $P^{\dag}=p(x_1)p(x_2)p(\mathbf{x}_r)p(\hat{y}_{r1})p(\hat{y}_{r2})$ where $p(\mathbf{x}_r)=\prod_{k=1}^{N}p(x_{rk})$. Second, for each block $j\in\{1:B\}$ and according to $P^{\dag}$, randomly and independently generate $2^{nR_\mu}$ codewords $x_\mu^n(w_{\mu,j})$ that encode $w_{\mu,j}$ where $\mu\in\{1,2\}$. Third, similarly generate $2^{nR_{r1}}$ $(2^{nR_{r1}})$ codewords $\hat{y}_{r1,j}^n(l_{1,j})$ $(\hat{y}_{r2,j}^n(l_{2,j}))$ that encode $l_{1,j}$ $(l_{2,j})$, where $R_{r1}$ $(R_{r2})$ is the transmission rate of $l_{1,j}$ $(l_{2,j})$ by ${\cal R}$. Last, for each pair $(l_{1,j-1},l_{2,j-1})$, generate $2^{n(R_{r1}+R_{r2})}$ codewords $\mathbf{x}_r^n(l_{1,j-1},l_{2,j-1})$.
\subsubsection{Encoding}
Let $(w_{1,j},w_{2,j})$ be the messages to be sent in block $j$.  Then, $\text{UE}_{k}$ transmits $x_k^n(w_{k,j})$. Moreover, since ${\cal R}$ has already estimated $(\hat{l}_{1,j-1},\hat{l}_{2,j-1})$ in block  $j-1$, it  transmits $\mathbf{x}_r^n(\hat{l}_{1,j-1},\hat{2}_{1,j-1})$ in block $j$. 
${\cal R}$ also
 utilizes $\tilde{y}_{r1}^n(j)$ and $\tilde{y}_{r2}^n(j))$  to find $l_{1,j}$ and $l_{2,j}$  such that:
\begin{align}
\left(\hat{y}_{r1}^n(l_{1,j}),\tilde{y}_{r1}^n(j)\right)\in A_{\epsilon}^n, \; \text{and}
\left(\hat{y}_{r2}^n(l_{2,j}),\tilde{y}_{r2}^n(j)\right)\in A_{\epsilon}^n.
\end{align}
\noindent
By the Covering lemma \cite[Lemma $3.3$]{hsce612}, such $l_{1,j}$ and $l_{2,j}$ exist if:
\begin{align}\label{somin}
\!\!\!\!\!R_{r1}&>I(\hat{Y}_{r1};\tilde{Y}_{r1})\triangleq \zeta_1\;\text{and}\;
R_{r2}>I(\hat{Y}_{r2};\tilde{Y}_{r2})\triangleq \zeta_2.
\end{align}
\subsubsection{Decoding}
Without loss of generality, assume all transmitted messages and quantization indices are equal to $1$. Then,
${\cal D}$  utilizes $\tilde{y}_{d1}^n$ and $\tilde{y}_{d2}^n$  in block $j$ and $\tilde{\mathbf{y}}_{dr}$ in block $j+1$ to find a unique message pair  $(\tilde{w}_{1,j},\tilde{w}_{2,j})$ for some quantization index pair $(\tilde{l}_{1,j},\tilde{l}_{2,j})$ such that:
\begin{align}\label{JTR}
&\big(x_1^n(\tilde{w}_{1,j}),\hat{y}_{r1}^n(\tilde{l}_{1,j}),\tilde{y}_{d1}^n(j)\big)\in A_{\epsilon}^n,\;
\big(\mathbf{x}_r^n(\tilde{l}_{1,j},\tilde{l}_{2,j}),\tilde{\mathbf{y}}_{dr}^n(j+1)\big)\in A_{\epsilon}^n,\nonumber\\
&\text{and}\;\;\big(x_2^n(\tilde{w}_{2,j}),\hat{y}_{r2}^n(\tilde{l}_{2,j}),\tilde{y}_{d2}^n(j)\big)\in A_{\epsilon}^n.
\end{align}
Let $J_1,$ $J_2$ and $J_3$ be as follows:
\begin{align}\label{JJJ}
\!\!\!\!J_k\triangleq& I(\hat{Y}_{rk};X_k,\tilde{Y}_{dk})+I(\mathbf{X}_r;\tilde{\mathbf{Y}}_{dr}),\;
J_3\triangleq I(\hat{Y}_{r1};X_1,\tilde{Y}_{d1})+I(\hat{Y}_{r2};X_2,\tilde{Y}_{d2})+I(\mathbf{X}_r;\tilde{\mathbf{Y}}_{dr}),
\end{align}
for $k\in\{1,2\}$. Then, applying JT analysis \cite{hsce612} to (\ref{JTR}) leads to some rate constraints, which along with   (\ref{somin}) leads to the following constraints: 
\begin{align}\label{mdec2}
\!\!\!\!R_k\leq& I(X_k;\hat{Y}_{rk},\tilde{Y}_{dk})\triangleq J_4,\;R_k\leq I(X_k;\tilde{Y}_{dk})+J_k-\zeta_k\triangleq J_5,\;k\in\{1,2\}\\
\!\!\!\!R_k\leq& I(X_k;\tilde{Y}_{dk})+J_3-\zeta_1-\zeta_2\triangleq J_6,\;\text{and}\;
R_1+R_2\leq I(X_1;\tilde{Y}_{d1})+I(X_2;\tilde{Y}_{d2})+J_3-\zeta_1-\zeta_2\triangleq J_7.\nonumber
\end{align}
In (\ref{mdec2}), the constraint with $J_5$ is redundant since $J_6<J_5$. We obtain $I_1, \ldots, I_5$ in (\ref{sth1rr1}) by applying  (\ref{mdec2}) to the ZF-processed received signals in (\ref{gaumodzf}) with $i)$ the signaling in (\ref{sigG}), $ii)$ the approximations in \cite{lars} where for $k\in\{1,2\},$
\begin{align}
\!\!\!\!\!\norm{\boldsymbol{a}_{rk}}^2&\rightarrow d_{rk}^{-\alpha}(N-2)\approx d_{rk}^{-\alpha}N,\;
\norm{\boldsymbol{a}_{dv}}^2\rightarrow d_{dv}^{-\alpha}(M-N-2)\approx d_{dv}^{-\alpha}(M-N)\;v\in\{1,2,r\},\nonumber
\end{align}
and $iii)$ the equivalency of the rate $(I(\mathbf{X}_r;\tilde{\mathbf{Y}}_{dr}))$ from ${\cal R}$ to ${\cal D}$ to that from $N$ single-antenna UEs to ${\cal D}$ since $M\gg N$ \cite{lars}.
\vspace*{-4mm}
\section*{Appendix B: KKT Conditions for Theorem \ref{cor1K}}
Considering Step $5$ 
of the Theorem \ref{cor1K} proof, the Lagrangian function 
of (\ref{newop}) is given as:
\begin{align}\label{LAGOP}
  L(\overline{\delta},\eta_s,\overline{\eta})=& \sum_{k=1}^{K}\mu_k \ln\left(a_k+\frac{b_k}{e^{\delta_k}}\right)-\sum_{k=1}^{K}\eta_k\delta_k
  +\eta_s\Big(-\ln(\lambda_s)+\sum_{k=1}^{K}\delta_k\Big),
\end{align}
where $\boldsymbol{\delta}=[\delta_1\;\ldots\;\delta_K]$ is the optimal quantization parameter vector, $\eta_s$ is the Lagrangian multiplier associated with the equality constrains, and $\boldsymbol{\eta}=[\eta_1\;\ldots\;\eta_K]$ is the
Lagrangian multiplier vector associated with the inequality constraints. The KKT conditions are:
\begin{subequations}\label{grp}
\begin{align}
&\nabla_{\overline{\delta}}L(\overline{\delta},\eta_s,\overline{\eta})=0
\Rightarrow\;\frac{\mu_kb_k}{a_ke^{\delta_k}+b_k}+\eta_k=\eta_s,\quad\eta_s\geq 0,\;-\delta_k\leq0,\; \eta_k\geq 0,\; \eta_k\delta_k=0\label{FR}\\
&-\ln(\lambda_s)+\sum_{k=1}^{K}\delta_k=0,\label{Sec}
\end{align}
\end{subequations}
where $\nabla_x f$ is the gradient of $f(\cdot)$ with respect to $x$. Note that to jointly satisfy the conditions in (\ref{FR}), we consider the following two cases for the optimal $\boldsymbol{\delta}$:
\begin{itemize}[leftmargin=*]
  \item If $\delta_k>0,$ then $\eta_k=0$ and
  \begin{align}\label{cas1}
  \frac{\mu_kb_k}{a_ke^{\delta_k}+b_k}=\eta_s\Rightarrow \delta_k=\ln\Big(\frac{\mu_kb_k}{a_k}(\eta_s^{-1}-\mu_k^{-1})\Big).
  \end{align}
  Furthermore, since $e^{\delta_k}>1$, we have $\eta_s<\omega_k^{-1}\Leftrightarrow x_s>\omega_k,$ where $x_s=\eta_s^{-1}$.
  \item If $\delta_k\leq0,$ then $\eta_k>0$ and $\omega_k^{-1}+\eta_k=\eta_s\Rightarrow \eta_s> \omega_k,\Leftrightarrow x_s< \omega_k$.
\end{itemize}
From the above two cases, we have:
\begin{align}\label{xs_o}
\delta_k^{\ast}=\ln\big(\frac{\mu_kb_k}{a_k}(x_s^{\ast}-\mu_k^{-1}))\big)\; \text{if}\;x_s^{\ast}> \omega_k,\;
\text{or}\; \delta_k^{\ast}=0\;\text{if}\;x_s^{\ast}\leq \omega_k.
\end{align}
Next, to determine $x_s^{\ast}$, we first arrange $\omega_k$'s in an increasing order as in (\ref{order}). Then,
considering the condition (\ref{Sec}), we determine the maximum $\upsilon^{\ast}\in\{1,2,\ldots,K\}$ such that $x_s>\omega_k$ for $k\leq\upsilon^{\ast}$ while $x_s\leq \omega_k$ for $k>\upsilon^{\ast}$. Hence, by (\ref{xs_o}), the condition (\ref{Sec}) is expressed as follows:
\begin{align}\label{chqr}
&\sum_{k=1}^{\upsilon^{\ast}}\delta_k-\ln(\lambda_s)=0,\Leftrightarrow \prod_{k=1}^{\upsilon^{\ast}}\mu_kb_ka_k^{-1}(x_s-\mu_k^{-1})-\lambda_s=0,\nonumber\\
&\Leftrightarrow 0=\prod_{k=1}^{\upsilon^{\ast}}(x_s-\mu_k^{-1})-\lambda_s\prod_{k=1}^{\upsilon^{\ast}}(\mu_kb_k)^{-1}a_k\triangleq f(x_s).
\end{align}
To determine $\upsilon^{\ast}$, $f(x_s)$ should have a real positive root that is greater than $\omega_{\upsilon^{\ast}}$, i.e., $x_s>\omega_{\upsilon^{\ast}}$. We check the existence of such a root using the  Descartes' rule of signs where the number of positive real roots in a polynomial is equal to  either the number of sign changes between consecutive nonzero coefficients, or is less than that by an even number.
Consequently,  $f(x_s)$ has a root $x_s>\omega_{\upsilon^{\ast}}$ if  $f(x_s+\omega_{\upsilon^{\ast}})$ has  a real positive root. From (\ref{chqr}),
\begin{align}
f(x_s+\omega_{\upsilon^{\ast}})=\prod_{k=1}^{\upsilon^{\ast}}
(x_s+\omega_{\upsilon^{\ast}}-\mu_k^{-1})-\lambda_s\prod_{k=1}^{\upsilon^{\ast}}(\mu_kb_k)^{-1}a_k.\nonumber
\end{align}
Since $\omega_{\upsilon^{\ast}}>\omega_k$ for $k<\upsilon^{\ast}$ as in (\ref{order}), we have $\omega_{\upsilon^{\ast}}-\mu_k^{-1}>0$. Therefore, all nonzero coefficients of $f(x_s+\omega_{\upsilon^{\ast}})$ are positive except the last constant term that is given as follows:
\begin{align}
\prod_{k=1}^{\upsilon^{\ast}}(\omega_{\upsilon^{\ast}}-\mu_k^{-1})-\lambda_s\prod_{k=1}^{\upsilon^{\ast}}
(\mu_kb_k)^{-1}a_k.
\end{align}
If this constant is negative, we have one sign change in $f(x_s+\omega_{\upsilon^{\ast}})$, i.e., one positive real root. Otherwise, the maximum positive real root is less than $\omega_{\upsilon^{\ast}}$. Using this criterion, we can determine $\upsilon^{\ast},$ $x_s^{\ast},$ $\delta_k^{\ast},$ and $\lambda_k^{\ast}$ as in Theorem \ref{cor1K}.
\vspace*{-6mm}
\bibliographystyle{IEEEtran}
\bibliography{references}
\end{document}